\documentclass[fleqn,10pt]{wlscirep}
\usepackage[utf8]{inputenc}
\usepackage[T1]{fontenc}
\usepackage{tabularx}

\myexternaldocument{SI}

\usepackage{float}
\usepackage{setspace}
\usepackage[skip=3pt plus1pt, indent=20pt]{parskip}
\usepackage{blindtext}
\begin{document}

\vspace{2cm}
{\LARGE \textbf{Liquid-activated quantum emission from pristine hexagonal\\ 
\vspace{0.5cm} boron nitride for nanofluidic sensing}}\\

\noindent
{Nathan Ronceray}\textsuperscript{1,2,*}, {Yi You}\textsuperscript{3,4}, {Evgenii Glushkov}\textsuperscript{1}, {Martina Lihter}\textsuperscript{1,5}, {Benjamin Rehl}\textsuperscript{2}, {Tzu-Heng Chen}\textsuperscript{1}, {Gwang-Hyeon Nam}\textsuperscript{3,4}, {Fanny Borza}\textsuperscript{1}, {Kenji Watanabe}\textsuperscript{6}, {Takashi Taniguchi}\textsuperscript{7}, {Sylvie Roke}\textsuperscript{2}, 
{Ashok Keerthi}\textsuperscript{4,8}, {Jean Comtet}\textsuperscript{9}, {Boya Radha}\textsuperscript{3,4,*}, {Aleksandra Radenovic}\textsuperscript{1,*}
\vspace{0.5cm}

\noindent
{\small
1. {Laboratory of Nanoscale Biology, Institute of Bioengineering (IBI), School of Engineering (STI), École Polytechnique Fédérale de Lausanne (EPFL), Lausanne, Switzerland. }\\
2. {Laboratory for Fundamental BioPhotonics, Institute of Bioengineering (IBI), School of Engineering (STI), École Polytechnique Fédérale de Lausanne (EPFL), Lausanne, Switzerland.}\\
3. {Department of Physics and Astronomy, School of Natural Sciences, The University of 
Manchester, Manchester, United Kingdom}\\
4. {National Graphene Institute, The University of Manchester, Manchester, United Kingdom}\\
5. {Present address: Institute of Physics, HR-10000 Zagreb, Croatia}\\
6. {Research Center for Electronic and Optical Materials, National Institute for Materials Science, 1-1 Namiki, Tsukuba 305-0044, Japan}\\
7. {Research Center for Materials Nanoarchitectonics, National Institute for Materials Science,  1-1 Namiki, Tsukuba 305-0044, Japan}\\
8. {Department of Chemistry, School of Natural Sciences, The University of Manchester, Manchester, UK}\\
9. {Soft Matter Sciences and Engineering, ESPCI Paris, PSL University, CNRS, Sorbonne Université, 75005 Paris, France\\}
}\\
\vspace{0.5cm}
{\small nathan.ronceray@epfl.ch, radha.boya@manchester.ac.uk, aleksandra.radenovic@epfl.ch}

\vspace{0.5cm}

\section*{Abstract}
\hspace{0.8 cm}Liquids confined down to the atomic scale can show radically new properties. However, only indirect and ensemble measurements operate in such extreme confinement, calling for novel optical approaches enabling direct imaging at the molecular level. Here, we harness fluorescence originating from single-photon emitters at the surface of hexagonal boron nitride (hBN) for molecular imaging and sensing in nanometrically confined liquids. The emission originates from the chemisorption of organic solvent molecules onto native surface defects, revealing single-molecule dynamics at the interface through spatially correlated activation of neighboring defects. Emitter spectra further offer a direct readout of local dielectric properties, unveiling increasing dielectric order under nanometer-scale confinement. Liquid-activated native hBN defects bridge the gap between solid-state nanophotonics and nanofluidics, opening new avenues for nanoscale sensing and optofluidics. 
\vspace{1 cm}
\section*{Main Text} 
\hspace{0.8 cm}Nanostructures made of two-dimensional (2D) materials have become prominent in nanofluidic research\cite{faucher2019critical,bocquet2020nanofluidics}. Liquid confinement to a few molecular layers between atomically smooth walls has led to anomalies in molecular transport\cite{majumder2005enhanced, radha2016molecular,secchi2016massive, mouterde2019molecular} and structure\cite{agrawal2017observation,fumagalli2018anomalously}. Nevertheless, the direct observation of these emerging phenomena remains challenging due to limitations of current techniques in extreme confinements, where even molecular fluorophores cannot penetrate\cite{grimm2022caveat}. This calls for the development of imaging methods that can access molecular properties in confinement\cite{faucher2019critical,bocquet2020nanofluidics}.

To tackle this goal, solid-state optically active defects show promise. Fluorescent defects in diamond have enabled optical probing of nanoscale matter\cite{schirhagl2014nitrogen}, including liquids\cite{aslam2017nanoscale}, but they cannot be easily embedded in 2D nanostructures. Coincidentally, hexagonal boron nitride (hBN) has been used in both nanofluidics\cite{siria2013giant,fumagalli2018anomalously,mouterde2019molecular} and nanophotonics, where various point defects within its 6 eV band gap have been identified as room-temperature quantum emitters\cite{tran2016quantum, hayee2020revealing,kianinia2022quantum}. While emitters in hBN have been induced artificially using techniques such as irradiation\cite{choi2016engineering, fournier2021position} or carbon doping\cite{mendelson2021identifying,koperski2020midgap}, the potential of liquid treatments remains largely unexplored. Recent studies have combined liquid and irradiation treatments to activate plasma-induced surface defects in hBN using water\cite{comtet2020direct} and binary mixtures of water with organic solvents\cite{comtet2021anomalous}. Post-treatment of ion beam-exposed hBN with liquids has also been shown to modify defect emission properties\cite{glushkov2022engineering}. Furthermore, plasma-induced surface defects have been utilized to study single interfacial charge dynamics, such as proton hopping\cite{comtet2020direct,comtet2021anomalous}. However, even mild plasma treatment induces mechanical and chemical changes that result in hBN crystals no longer having atomically smooth surfaces\cite{na2021modulation}, thus preventing the integration of defects in ultra-flat van der Waals heterostructures, which are crucial for advancements in angstrom-scale fluidics\cite{radha2016molecular,fumagalli2018anomalously}.

In this study, we demonstrate that organic solvents can activate visible-range quantum emission from pristine high-quality hBN crystals. We attribute this phenomenon to the interaction between organic molecules and native surface defects\cite{taniguchi2007synthesis,wong2015characterization,schue2016characterization,henck2017direct}. By employing spectral super-resolution microscopy\cite{comtet2019wide}, we observe defect-mediated molecular random walks and couplings between defect dipoles and the liquid medium, leading to tunable emission wavelengths through the dielectric properties of the liquid. Leveraging intrinsic properties of widely used 2D materials and common solvents, the fluorescence activation mechanism reported here is utilized to image nanofluidic structures, with emitters serving as nanoscale probes of the order and dynamics of liquid media confined to the nanoscale.

\subsection*{Liquid-activated fluorescence from pristine hBN}
\hspace{0.8 cm}The surface of untreated hBN crystals exhibits visible-range fluorescence when in contact with common organic solvents like ethanol. To demonstrate this effect, we exfoliated high-quality hBN crystals\cite{taniguchi2007synthesis} onto a glass coverslip, which was placed in a liquid-filled chamber on an inverted microscope (Fig. 1a). Photoluminescence (PL) of the crystals under 561 nm wide-field laser illumination (0.35-3.5 kW/cm$^2$) was collected using a high-numerical aperture objective and projected onto a camera chip. Pristine hBN in air or water did not exhibit fluorescence under these illumination conditions. However, we observed intense fluorescence from as-exfoliated crystals in contact with ethanol (Fig. 1b). The fluorescence intensity gradually decreased over continuous illumination and stabilized after several seconds, revealing emission from sub-diffraction spots (Fig. 1c and Supplementary Video 1) which can be localized with $\approx$10nm precision using single-molecule localization microscopy\cite{feng2018imaging} (details in Methods). We attribute this emission to  the activation of defects present in the as-exfoliated crystal through contact with the liquid, resulting in randomly distributed transient emitters on the surface (Supplementary Fig. 1). Under constant 3.5 kW/cm$^2$ illumination, the number of emitters decreased to a stable value of approximately 0.4 per square micron (Fig. 1d). This process does not deteriorate the crystal (Supplementary Fig. 2 \& 3), and the decrease observed in Fig. 1c is reversible: when left in the dark, the crystal fluorescence recovered within tens of minutes (Supplementary Fig. 4) without inducing additional emitters in the steady state (Supplementary Fig. 5).

Remarkably, the fluorescent activation of the surface occurred with most common organic solvents, including n-alkanes (pentane to hexadecane) and primary alcohols (methanol to 1-pentanol). However, we did not observe any emission in pure water, heavy water, and hydrogen peroxide. To quantitatively compare the steady-state fluorescence in different liquid media, we imaged freshly cleaved hBN crystals in several liquids under 3.5 kW/cm$^2$ illumination. The steady-state fluorescence can be quantified using the crystal brightness $I_\text{crystal}$, defined as the sum of localized emitter intensities per surface unit and time unit (Fig. 1e). From these observations, we can classify solvents into three types based on the extent of hBN fluorescent activation. Most organic solvents, such as primary alcohols, n-alkanes, and chloroalkanes, exhibited intense fluorescence ($I_{\text{crystal}}>500$ photons/µm$^2$/s), representing type I activation. Glycerol and other high-boiling point liquids ($\geq 200^\circ C$) exhibited a limited but measurable level of fluorescence, classified as type II. On the other hand, pure water showed no activation, falling into type III ($I_{\text{crystal}}<10$ photons/µm$^2$/s). Furthermore, the addition of 10v/v\% of water in ethanol reduced the number of emitters, resulting in a ~5-fold reduction in the fluorescence signal (Supplementary Fig. 6). The distinction between activation types could not be solely explained by physical parameters of the solvents (Supplementary Fig. 7), indicating a chemical specificity. The steady-state emitter density was found to depend on the liquid medium (Supplementary Fig. 7 \& 8), the illumination power (Supplementary Fig. 9), and macroscopic flow over the crystal (Supplementary Fig. 10).

\begin{figure}[H]
\centering
\includegraphics{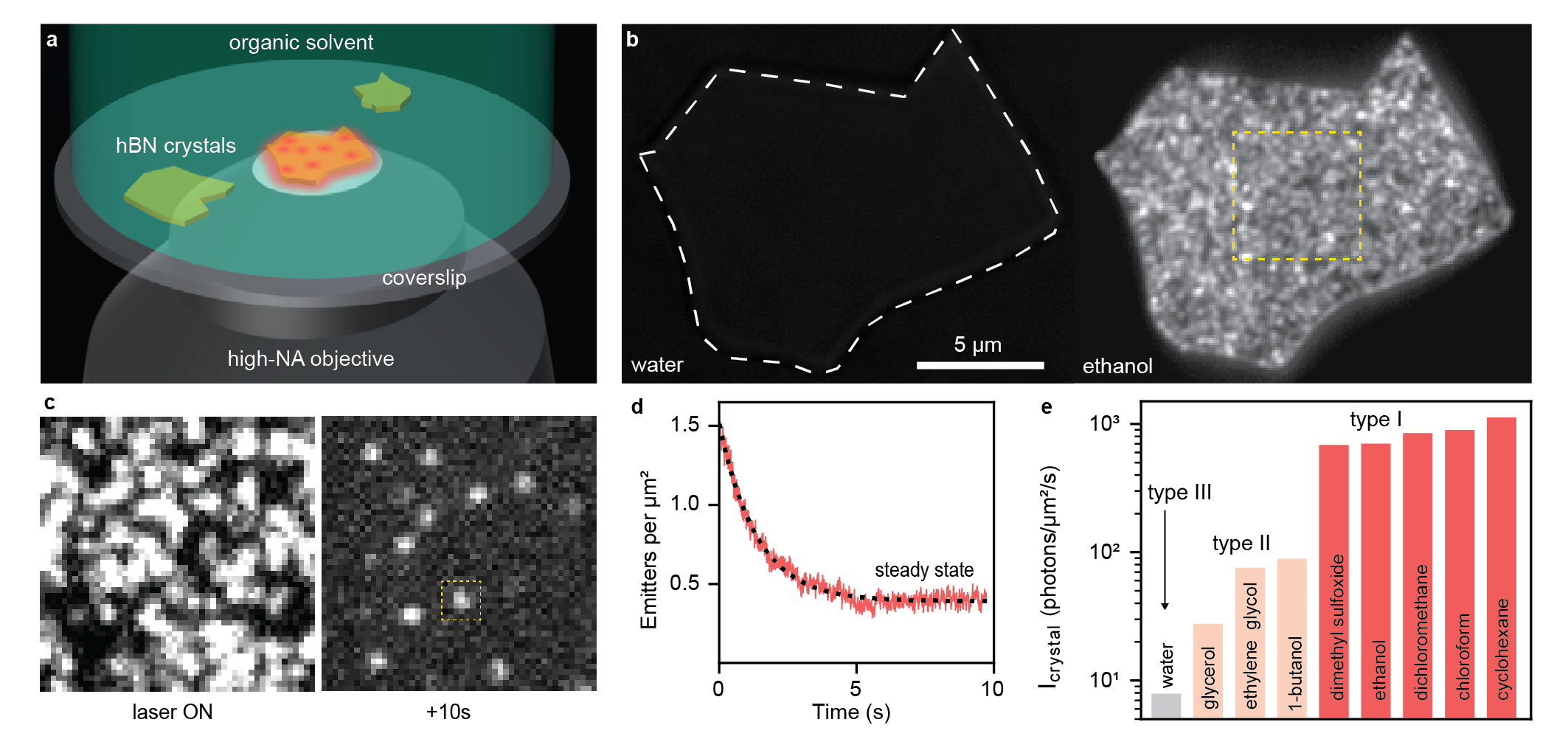}
\caption{\textbf{Liquid-induced fluorescence from pristine hBN crystals}. 
\textbf{a}, Sketch of the experimental setup. \textbf{b}, Wide-field fluorescence images of a hBN crystal under 3.5 kW/cm$^2$ 561 nm laser light illumination with 1 second exposure time. No fluorescence was observed in water, but in ethanol the entire crystal surface became fluorescent. The images underwent linear contrast enhancement. \textbf{c}, A zoomed-in view of the dashed yellow box in \textbf{b} reveals dense clusters of emission when turning the laser is turned ON, with 6 ms exposure time. After 10 s of wide-field illumination, the crystal surface reached a stable number of diffraction-limited isolated emitters. \textbf{d}, Localization microscopy-based counting of the emitters as a function of the illumination time: after 5 seconds, a steady state was reached. The dashed line is a fit to an offset exponential relaxation. \textbf{e}, Liquid dependency of the crystal fluorescence, showing strongly activating liquids (type I), mildly activating liquids (type II) and no activation in water (type III).}
\label{fig:fig1}
\end{figure}

Considering the dynamics of these fluorescent emitters, a striking observation is the presence of fluorescent trajectories on the crystal surface (Supplementary Movie 2). These trajectories indicate the correlated activation of neighboring defects, corresponding to molecular random walks. By linking the super-resolved localizations of emitters (see Methods), we can extract the associated trajectories, as illustrated in Fig. 2b. 
Previously observed trajectories on plasma-exposed hBN in water and binary mixtures of water and alcohols were attributed to proton hopping\cite{comtet2020direct,comtet2021anomalous}. A similar phenomenology is thus expected here, with (i) defect activation due to reversible charge transfer from and to the solvent and (ii) correlated activation of neighboring defects, mediated by the lateral motion of charge-bearing solvent molecules that remain physisorbed on the crystal surface. However, the aprotic nature of some solvents used here, and the spectral differences point to a distinct emitter type and reactivity. Since neither pristine hBN nor the liquids used possess visible-range electronic transitions, the activation of emitters at the interface must arise from an electronic structure rearrangement that generates this new optically addressable electronic transition. This could be explained by the chemisorption of organic molecules onto hBN defects\cite{lvova2016theoretical}. The chemical selectivity observed in Figure 1e suggests that a necessary and sufficient condition for a pure liquid to activate native hBN defects is the presence of a carbon atom in its molecular structure. This finding aligns with recent research indicating the crucial role of carbon in activating visible-range emitters in hBN\cite{mendelson2021identifying}. While a direct observation of the exact chemical structure of the emitters is challenging, the physicochemical interactions between the defects and the liquid, along with the photophysical properties of the emitters will guide structural assignment.

\subsection*{Analysis of emitter dynamics}
\hspace{0.8 cm}To understand the trajectories quantitatively, we acquired 50k, 6ms-long frames of a 13$\times$13µm crystal area in the steady state, yielding 700k localizations leading to 100k trajectories. Figure 2a shows a subset of these trajectories overlaid with the super-resolved image obtained from 5000 frames. This visualization reveals that while some trajectories exhibit free hopping behavior, others remain trapped for extended periods, resulting in bright spots in the super-resolved image. Most emitters are active for tens to hundreds of milliseconds, but a fraction displays stability for several seconds (Fig. 2c), with some activations lasting over a minute (Supplementary Fig. 11). As  measurements were conducted at a low emitter density where trajectories do not split or merge with statistical significance, we assign trajectories to single molecules binding to single defects.

We analyzed the observed molecular random walks through their trajectory residence times on the crystal surface $T_\text{res}^T$, which comprise complex information on both the chemisorption energy at defect sites and the physisorption energy on pristine hBN in between defects, as well as residence times of molecules at single defect sites $T_\text{res}^D$ corresponding to chemisorption only. Both residence times were found to follow a double exponential decay (Fig. 2d), with slow exponential decay components $\tau_\text{res}^D=$35$\pm$1 ms and $\tau_\text{res}^T=$82$\pm$3 ms, respectively. Assuming that single-defect residence times follow the Arrhenius equation $\tau_\text{res}^D=\nu^{-1} e^{\Delta G/kT}$ where $\nu \approx 10^{12}-10^{13} \, s^{-1}$ is a molecular attempt rate\cite{comtet2021anomalous}, we obtain a desorption energy barrier $\Delta G \approx$ 24-27 kT$\approx$ 0.6-0.7 eV which is larger than typical physisorption energies (tens of meV) and smaller than covalent bonding energies (several eV)\cite{huber2019chemical}. This can be rationalized in terms of a lowered energy barrier under illumination\cite{comtet2020direct}consistent with the observed light-induced reduction in number of emitters in Fig. 1d as well as the illumination power dependency of the density of emitters (Supplementary Fig. 9).

\begin{figure}[H]
\centering
\includegraphics{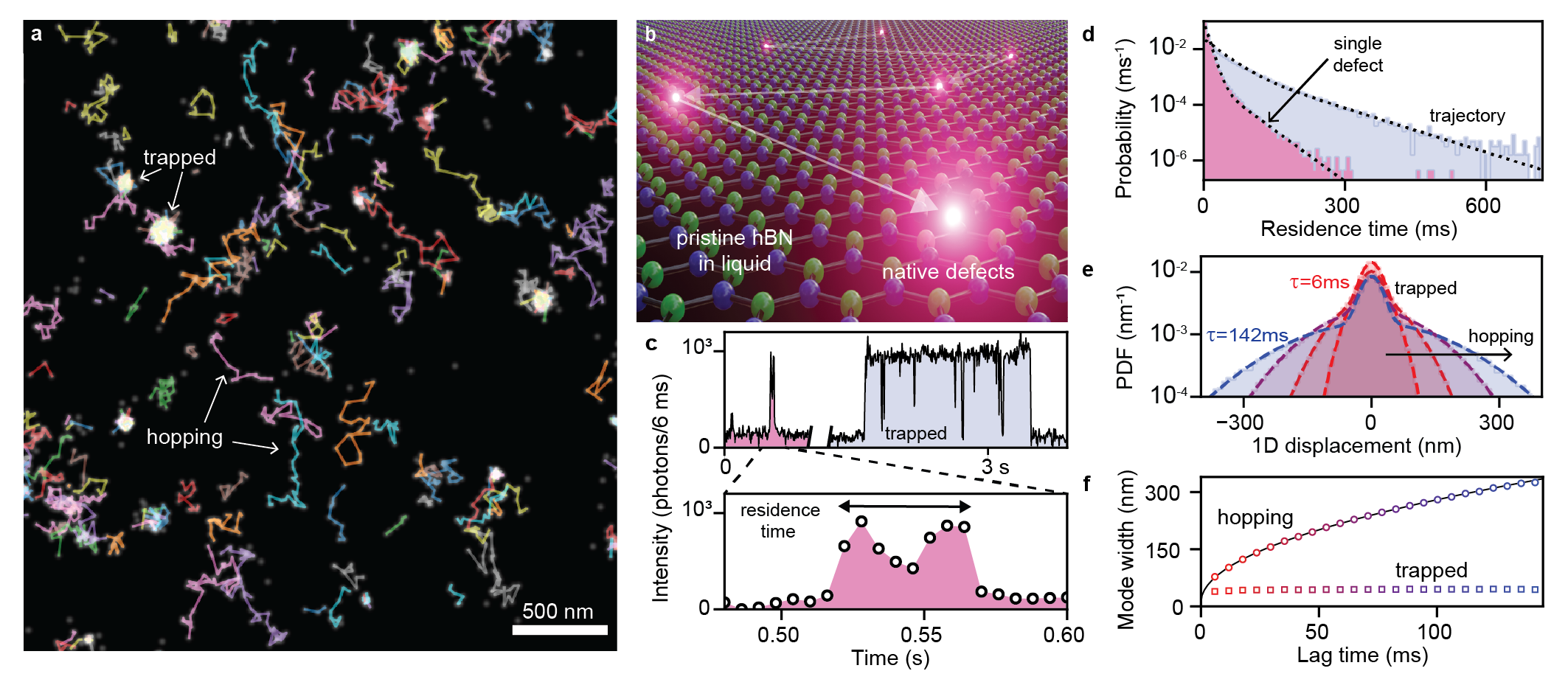}
\caption{\textbf{The surface of pristine hBN reveals interfacial molecular dynamics}. 
 \textbf{a}, Overlay of a super-resolved image from 5000 frames showing hopping emitters in isopropanol as the linked trajectories, as well as trapped spots.
 \textbf{b}, Artist's view of the correlated activation of neighboring defects leading to trajectories.
 \textbf{c}, Representative intensity traces from the same images, taken from 7x7 pixel bins around emitters as delimited by the dashed yellow box in Fig. 1c, with 6 ms exposure time. The top right trace corresponds to a long defect activation. The top left trace corresponds to a short activation of the same defect, magnified on the bottom panel. \textbf{d}, Distribution of residence times on single defects and for the entire trajectories. Dotted lines are fits to a two-component exponential decay. \textbf{e}, Displacement probability density functions (PDF) of trajectories after different lag times $\tau$=6,24,66,142 ms. Dashed lines are fits to two-component Gaussians. \textbf{f}, Visualizing the evolution of the two modes of the Gaussian fit in Fig. 2e with increasing lag time. The central region, corresponding to the trapped state, remains of constant width, whereas the tails, corresponding to hopping, enlarge with time. The solid line is a fit to a standard diffusion curve.}
\label{fig:fig2}
\end{figure}

Turning our attention to the emitter motion, we find that their one-dimensional displacement probability density function\cite{anthony2006methods} PDF(x,$\tau$) follows a two-component Gaussian distribution (Fig. 2e). The central part of the distribution remains of constant size ($\approx$20 nm) corresponding to the localization uncertainty when a trajectory is trapped. The tails of the distribution however enlarge with increasing lag time, characterizing hopping events. As shown in Figure 2f, the hopping tail size scales as $\sqrt{2 D \tau}$ with lag time $\tau$, which corresponds to Brownian diffusion with a diffusion coefficient $D $= 9.1$\times$10$^{-14}$m$^2$/s, which is over four orders of magnitude slower than bulk liquid molecular diffusion coefficients. While this observation makes it impossible to resolve the molecular travel time between defects, this slowdown enables the detection of bright emission from localized spots, which we now propose as a spectral sensing tool.

\subsection*{Spectral properties and solvatochromic sensing} 
\hspace{0.8 cm}We examined the spectral response of the emitters to their liquid environment using spectral single-molecule localization microscopy (sSMLM)\cite{comtet2019wide} which enables simultaneous localization and spectral characterization (Fig. 3a, details in Methods). Single-emitter spectra were found to be homogeneously distributed, with the appearance of a single population of emitters when exposed to the same liquid environment. Ensemble averaged spectra were consistently characterized by two peaks, classically attributed to the zero-phonon line (ZPL) and the phonon side band (PSB) for emitters embedded in a matrix (Fig. 3b).

Interestingly, these ensemble spectra appeared to depend strongly on the activating liquid, and more precisely on its static dielectric constant $\epsilon_\text{liq}$. We present in Figure 3b spectra of emitters obtained in the following liquids of increasing polarity: pentane, tert-butyl alcohol and DMSO. A notable polarity-induced solvatochromic redshift of the emission was gradually observed from nonpolar pentane (615 nm) to more polar tert-butanol (626 nm) to highly polar DMSO (641 nm). Beyond the ZPL shift, we observed changes in the PSB, which is less clearly defined for polar solvents. We report in Figure 3c the center wavelengths of both peaks as obtained from fitting to a two-Lorentzian model for several solvents ordered by increasing $\epsilon_\text{liq}$. We tested 1-pentanol, isopropanol and methanol on top of previously introduced liquids to interpolate dielectric constant values and found that both the ZPL and the PSB are redshifted by over 25 nm ($\approx$80 meV) in highly polar liquids compared with nonpolar alkanes. In the range $\epsilon_\text{liq}<25$, a linear dependency was observed between the ZPL wavelength and the dielectric constant, indicated by the dashed line (Fig. 3c) with a slope of approximately 1 nm per unit.

\begin{figure}[H]
\centering
\includegraphics{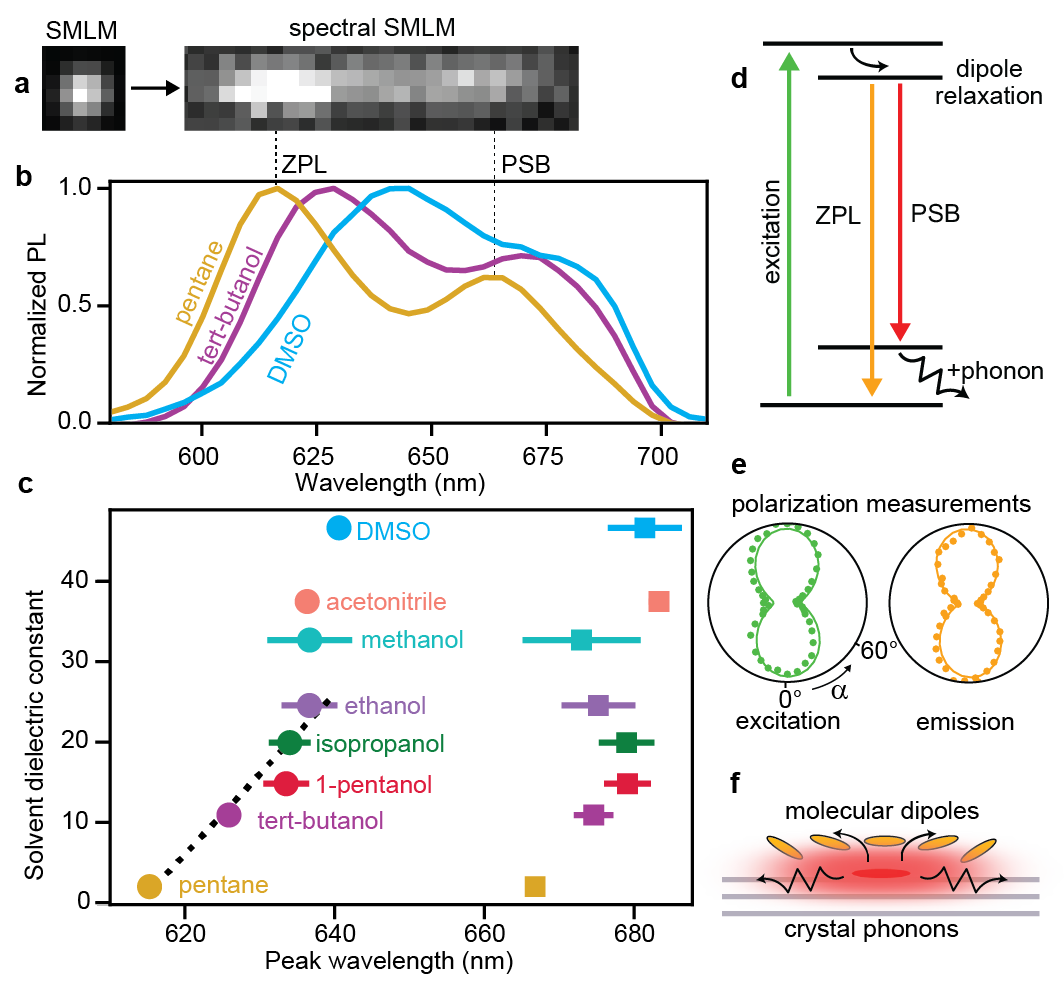}
\caption{ \textbf{Spectral properties of surface dipole emitters coupled to both solid and liquid environments}. \textbf{a}, Spectral single-molecule localization microscopy (sSMLM) splits the fluorescence signal from an emitter into a localization component (left, SMLM) and a spectral component (right) on the same camera chip. \textbf{b}, Ensemble spectra of liquid-activated emitters in different type I solvents exhibiting a clear zero-phonon line (ZPL) and phonon side band (PSB). \textbf{c}, Visualizing the wavelength shifts of both ZPL (circles) and PSB (squares), which correlate with the dielectric constant of the liquid. Peak positions were obtained by fitting to a sum of Lorentzians. The dashed line indicates the linear solvatochromic range, with a slope of 1 nm per unit. Error bars correspond to standard deviations of fitting parameters from groups of 100 single-molecule spectra. \textbf{d}, Jablonski diagram of processes at play: 561 nm laser excitation induces a dipolar excited state, which can emit directly (orange arrow, ZPL) or with emission of a phonon (red arrow, PSB).  \textbf{e}, Normalized intensity as a function of input or output light polarization angle $\alpha$ relative to the emitter axis. The solid lines correspond to fits to ideal electric dipole emission $\cos^2\alpha+\text{constant}$. More details are provided in Supplementary Figure 11. \textbf{f}, Sketch of an excited emitter that can interact with both the crystal through phonons and with surrounding molecules (yellow ellipses).}
\label{fig:fig3}
\end{figure}

 The Jablonski diagram presented in Figure 3d illustrates the process giving rise to the observed spectra. By absorbing a photon (excitation, green arrow), the emitter is excited to a dipolar state which interacts with the solvent and relaxes before radiating (dipole relaxation, curved arrow). Direct evidence of the dipolar nature of our liquid-activated emitters is presented in Figure 3e where either the linearly polarized light used for excitation (green) or the PL emission (orange) was rotated while monitoring the signal through an analyzer (details in Supplementary Fig. 11). As sketched in Figure 3f, the solvatochromic redshift can be described by the presence of liquid molecular dipoles stabilizing the excited state dipole, thus lowering its energy and redshifting the ZPL. After this step, as shown in Figure 3d, the transition back to the ground state can occur in two ways: direct emission of a photon (ZPL, orange arrow), and phonon-assisted emission (PSB, red arrow), which is redshifted compared to the ZPL as a fraction of the energy leads to lattice vibrations (phonon, zigzag arrow). Our analysis of the phonon side bands revealed increasing phonon broadening and decreasing phonon side band content with increasing solvent polarity (details in Supplementary Discussion and Supplementary Fig. 12). In the case of a mixture of polar ethanol and apolar heptane, the spectrum was very similar to that of the polar liquid, demonstrating a strong affinity between the emitters and polar molecules (Supplementary Fig. 13 \& 14).
\vspace{-0.3cm}
\subsection*{Time-resolved measurements reveal quantum emission}
\hspace{0.8 cm}To prove that the measured fluorescence originates from single-photon emitters, we performed time-correlated photon counting. For this, a 0.7 mW continuous-wave 561 nm laser beam was focused to a $\approx$ 1 µm$^2$ spot onto hBN crystals in liquid, and fluorescence signal was collected by two single photon detectors (SPDs) in a Hanbury Brown and Twiss interferometer configuration (inset of Fig. 4a). The typical time trace of a stable single emitter in acetonitrile is shown in Figure 4a. The analysis of photon arrival times from a 10s window shows a clear photon antibunching dip $g^{(2)}(0)=0.25\pm 0.02$ at zero delay time, demonstrating single-photon emission (Fig. 4b). This result implies that the bright spots are single emitters and not clusters, thereby their optical readout truly reports on nanoscale properties of the liquid. This activation of quantum emission through  the strong chemisorption interaction between a single activating molecule and a single defect was observed in carbon nanotubes\cite{he2017tunable}, but the mechanism at play here exhibits the particularities of being transient and observed in liquid. Photon statistics under pulsed excitation show the suppression of the correlated pulse peak at zero delay time, confirming single-photon emission (Fig. 4c). This feature was also found in hexadecane with a measured $g^{(2)}(0)=0.45\pm 0.04$. We thus demonstrated liquid-tunable single-photon emission with a ZPL shift of 21 nm (inset in Fig. 4b). This shift is comparable to those achieved by hBN defects in response to strain\cite{mendelson2020strain} or electric fields\cite{nikolay2019very}, which shows the potential of liquid-activated emitters as dielectric environment sensors. 

\begin{figure}[H]
\centering
\includegraphics[scale=0.96]{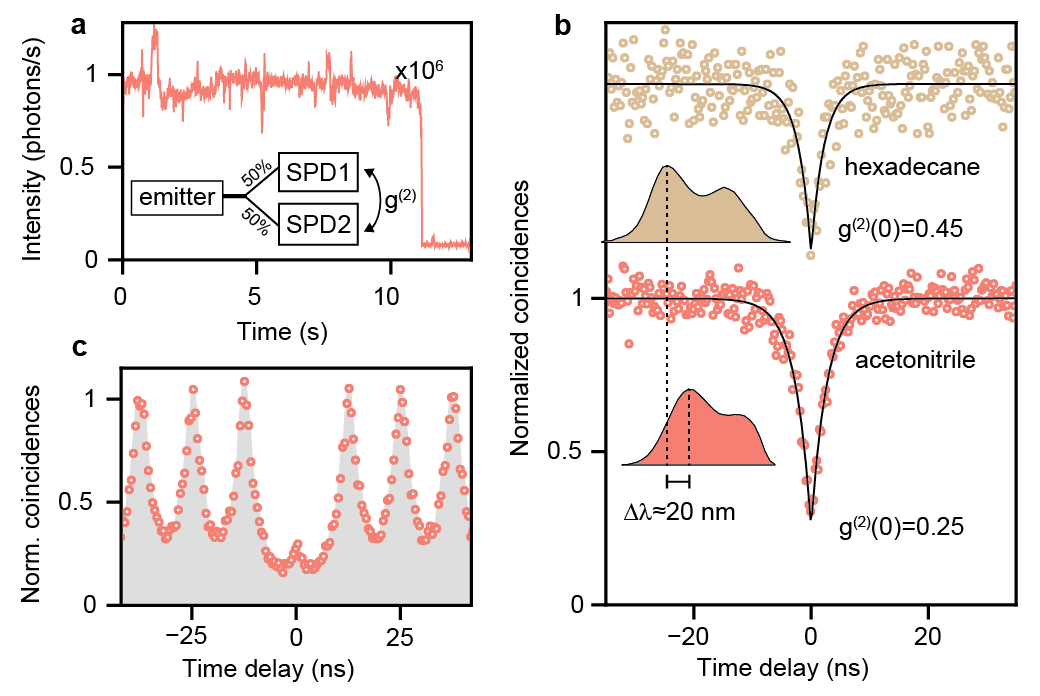}
\caption{\textbf{Quantum emission from liquid-activated emitters}. \textbf{a}, Representative PL trace from an isolated emitter in acetonitrile under 0.7 mW confocal excitation. The shaded region corresponds to the 10s-long trace used for photon statistics. \textbf{b}, Normalized coincidences $g^{(2)}$ measured from time correlated single photon counting in a Hanbury Brown and Twiss geometry, in hexadecane and acetonitrile. In both liquids, a pronounced antibunching is observed with $g^{(2)}(0)<0.5$ without background correction, proving the single-photon emission. The fluorescent lifetimes, corresponding to the width of the antibunching dip, were found to be 2.73$\pm$0.09 ns and 2.20 $\pm$ 0.20 ns for acetonitrile and hexadecane, respectively. Spectra are shown in the inset, demonstrating liquid-tunable single-photon emission from 615 nm to 636 nm. \textbf{c}, Single-photon statistics under pulsed laser excitation showing a suppression of the central peak due to antibunching.}
\label{fig:fig4}
\end{figure}

\subsection*{Integration in single-digit nanofluidic systems}
\hspace{0.8 cm}Building upon the characterization of the emission in bulk liquids, we probed hBN-liquid interfaces in molecular confinement in two-dimensional nanoslits. As sketched in Figure 5a, the nanoslits are obtained by van der Waals assembly of heterostructures comprising 3 crystals: bottom, spacer and top. The top crystal was chosen to be muscovite mica for its transparency and lack of fluorescent properties, and the bottom crystal was pristine hBN to be activated by the liquid.  The middle crystal, composed of few-layer graphene patterned by electron beam lithography acted as a spacer, defining a slit-shaped channel between the hBN and mica crystals, whose height was set by the number of 3.4 Å-thick graphene layers\cite{radha2016molecular} (details in Methods and in Supplementary Fig. 15).

An optical micrograph of a device with $h=2.4$ nm is provided in Figure 5c. The bottom blue region corresponds to the full heterostructure with nanoslits, and the top purple region corresponds to the open hBN crystal masked by graphene spacers but without encapsulation by the mica. We verified that covering the pristine hBN crystal with a patterned few-layer graphene crystal masks liquid-activated emitters, as was observed for other types of hBN emitters\cite{xu2020charge,stewart2021quantum}. On bare hBN, emitters are randomly distributed (Supplementary Fig. 16), but the graphene mask allows for their precise positioning on the basal plane of hBN in liquid. An overlay of the graphene spacer atomic force microscopy (AFM) topography and the super-resolved image is shown in Fig. 5b, demonstrating the correspondence between the lithographically defined graphene pattern and the optically measured fluorescence from masked hBN. We further verified that capping masked hBN with mica does not quench its fluorescence, allowing direct imaging of emitters in confined liquid (Fig. 5d). We show in Figure 5e that the localization intensity distributions with and without the confining mica top are similar, but the number of emitters in confinement is reduced by two thirds. Single-defect residence times inside nanoslits were found to be slightly longer in nanoslits than in masked hBN (Supplementary Fig. 17). Therefore, the observed three-fold decrease in emitter number comes not from faster photobleaching of the emitters but rather from confinement-induced slowdown of their activation kinetics.

Integrating emitters into nanofluidic structures allows probing the effect of confinement on liquid structure and dynamics. We first confirm the observation of trajectories in 1.4 nm confinement in Figure 5g, where a set of emitters is shown in the the 150 nm wide slit. We then focus on the spectral properties of confined emitters, which can be robustly extracted through sSMLM with relatively low numbers of localizations (<1000). We present sSMLM spectra obtained in the bare, masked, and confined geometries for ethanol (Fig. 5h) and acetonitrile (Fig. 5i). For both solvents, bringing the confinement size from 2.4 nm down to 1.4 nm leads to a clear blueshift pinpointed by the dashes indicating the ZPL position. For ethanol, the ZPL blueshifts from 637 nm to 624$\pm$2 nm, and for acetonitrile, from 636 nm to 621$\pm$1 nm under 1.4 nm confinement, bringing the spectral signature of a strongly polar solvent close to that of nonpolar alkanes. This substantial confinement-induced blueshift suggests that emitters experience a reduced dielectric constant of $\epsilon_\text{conf}\approx$11 in ethanol and 7 in acetonitrile when considering Figure 3c as a calibration curve.

Bringing the top mica wall close to the emitters can impact their emission in two ways. Firstly, confinement by the opposite wall of lowered dielectric constant $\epsilon_\text{wall}\approx 8$ could reduce the effective number of solvent molecules interacting with the emitter and confine the electric field lines within the slit\cite{kavokine2019ionic}, potentially destabilizing the excited state. Secondly, the emission can be affected by reducing the out-of-plane component of the liquid dielectric tensor\cite{fumagalli2018anomalously}. As depicted in Figure 5f, the interaction range between a solvent molecule with dipole $\mu_S$ and the defect with dipole $\mu_D$ is given by $\ell_\text{dip}=(\mu_S \mu_D/2\pi \epsilon_o k_B T)^{1/3}$. Assuming $\mu_D\sim a \times e$, where $a\approx 0.25$ nm is the in-plane lattice parameter of hBN and $e$ is the elementary charge, we estimate $\mu_D\approx$ 12 D. Using the dipole moments of ethanol (1.7 D) and acetonitrile (3.4 D), we find $\ell_\text{dip}\approx$ 1 nm and 1.3 nm, respectively, which are smaller than the height of the nanoslit. Hence, the observed effect under 1.4 nm confinement might not arise solely from geometrical effects due to the proximity of the top wall, and could be explained by the confinement-induced reduction of the solvent dielectric constant, as observed for water \cite{fumagalli2018anomalously} and predicted for other liquids\cite{motevaselian2020universal}. These results consolidate the picture of dipolar environment-tuned emitters, whose properties are affected by changes in the sensing hemisphere with volume $\sim 2/3 \pi  \ell_\text{dip}^3$, which encloses fewer than 100 molecules in the case of acetonitrile. Beyond passive diffusion and dielectric sensing of confined liquids, tracking emitter dynamics in confinement may be used to directly image nanoscale flow and study its interplay with defects\cite{seal2021modulating}.

\begin{figure}[H]
\centering
\includegraphics{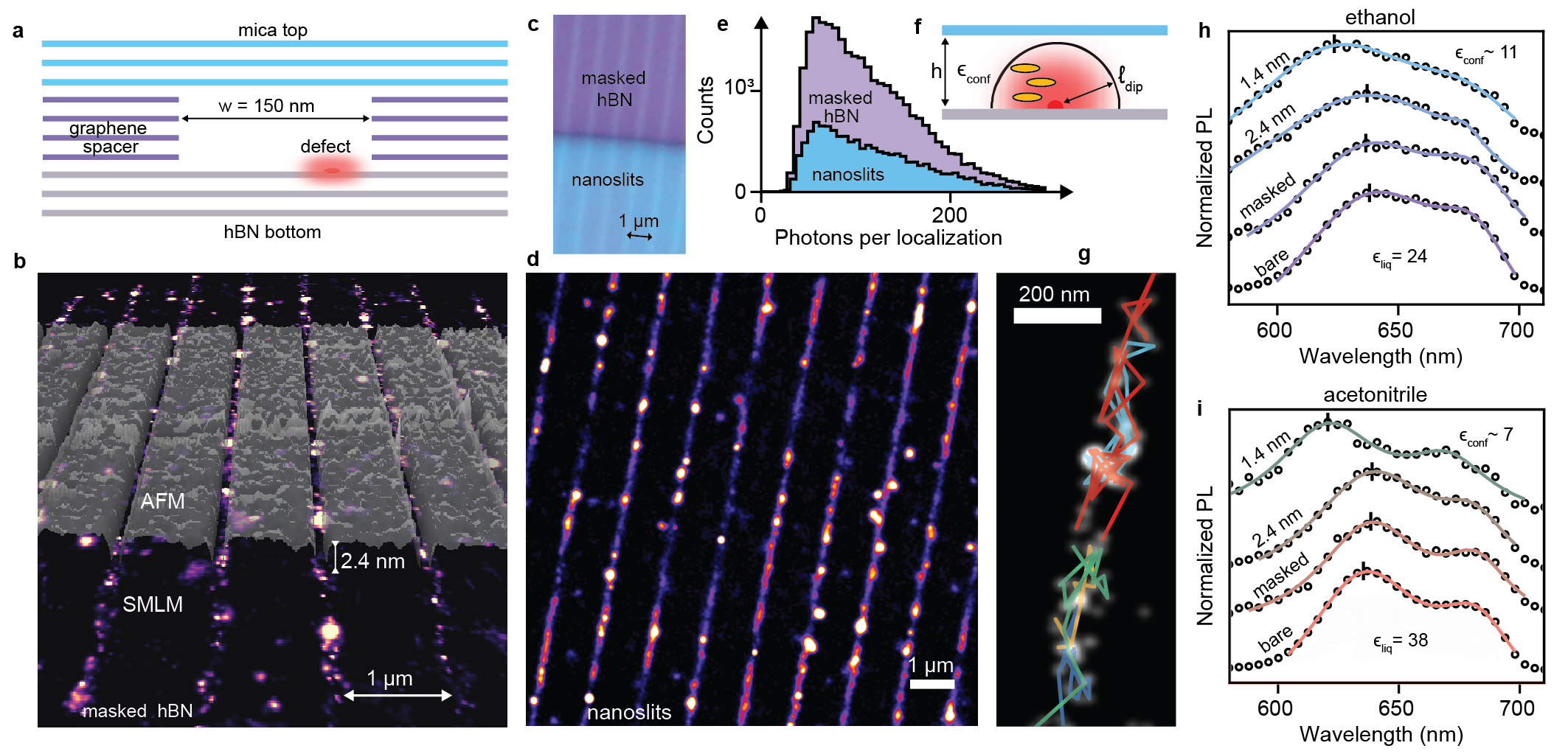}
\caption{\textbf{Nanoslit-embedded liquid-activated emitters}.  \textbf{a}, Sketch of the heterostructure nanoslit device. The red glow indicates an emitter inside the nanoslit. \textbf{b}, Overlay of a super-resolved image of masked ethanol-activated hBN and the AFM mapping of the graphene spacers. \textbf{c}, Optical micrograph of the heterostructure. On the purple colored part of the image, only graphene spacers on the hBN bottom crystal are present, which leads to masked hBN. The bottom blue region corresponds to the full heterostructure with slits. \textbf{d}, Super-resolved image of acetonitrile-activated emitters embedded in 2.4 nm-high nanoslits, from 30k frames with 20 ms exposure time and 1.4 kW/cm$^2$ illumination. \textbf{e}, Comparison of localization intensity distributions for masked hBN and 2.4 nm nanoslits in acetonitrile, showing no loss of photons but an overall reduction in number of localizations (details in Supplementary Fig. 18). \textbf{f}, Illustration of the effect of confinement: the liquid dielectric constant can be changed by confinement ($\epsilon_\text{conf}$) and the defect dipole can interact with solvent molecules (yellow ellipses) within a range $\ell_\text{dip}$, comparable to the confinement size $h$. \textbf{g}, Representative trajectories in 2.4 nm-high nanoslits filled with ethanol, overlaid with the super-resolved image. \textbf{h,i}, sSMLM spectra of liquid-activated defects in nanoslits filled with ethanol and acetonitrile, respectively. The confinement size is tuned from the open geometry to 2.4 nm down to 1.4 nm. Solid lines correspond to two-component Lorentzian fits, and black dashes indicate the extracted ZPL position.}
\label{fig:fig5}
\end{figure}

\subsection*{Outlook}

\hspace{0.8 cm}hBN crystals, already known for exceptional optical properties, exhibit a peculiar interaction with liquids. When in contact with organic solvents, native point defects on the atomically smooth surface of the crystal become emissive. This unique system, where the encounter of a single defect with a single organic molecule yields a single-photon emitter, combines solid-state emitters and organic fluorophores, providing a new tool for studying solid-liquid interfaces. We demonstrated two sensing approaches using liquid-activated hBN: the activation dynamics provide insights into interfacial charge transfer between defects and single molecules, while the emission spectra of the emitters offer information about the nanoscale dielectric environment. These phenomena were found to hold in confinement as small as a few nanometers, where only ensemble-averaged measurement techniques have been successful so far. As it relies on common samples and widely available single-molecule microscopy techniques, this approach could be readily applied for optical imaging and sensing in nanofluidic systems \textit{operando}.
\section*{Acknowledgements}

\hspace{0.7cm}

A.R., N.R., E.G., M.L., T.-H. C. acknowledge funding from the European Research Council (grant 101020445 - 2D-LIQUID), the Swiss National Science Foundation (grant 200021 192037) and the CCMX Materials Challenge grant “Large area growth of 2D materials for device integration”. B.Ra. acknowledges the funding from Royal Society University Research Fellowship URF/R1/180127. B.Ra., Y.Y., G-H.N acknowledge funding from the European Research Council (grant 852674 - AngstroCAP), and RS enhancement award RF/ERE/210016. A.K. acknowledges the funding from the Royal Society International Exchanges Award, IES/R1/201028 and EPSRC new horizons grant EP/V048112/1. K.W. and T.T. acknowledge support from the JSPS KAKENHI (Grant Numbers 20H00354, 21H05233 and 23H02052) and World Premier International Research Center Initiative (WPI), MEXT, Japan. J.C. acknowledges funding from the ANR (grant "GUACAmole" ANR-22-CE06-0003-01). S.R. thanks the Julia Jacobi fund.
We thank Edoardo Lopriore and Andras Kis for atomic force microscopy characterization; Aljoscha Söll and Zdeněk Sofer for providing hBN crystals grown with a different approach. N.R. acknowledges useful conversations with Marie-Laure Bocquet and Vasily Artemov. We thank Michael Weber and Miao Zhang for reporting independent checks of the activation phenomenon.

\section*{Author Contribution Statement}
N.R. observed the effect. A.R., B.Ra., J.C., A.K. and N.R. conceived the defect integration in nanoslits; A.R., N.R., M. L., T.-H. C., B.Ra., A.K. and J.C. designed experiments; E.G. fabricated the marked coverslips. Y. Y. and G.-H. N. fabricated the nanoslit devices under the supervision of B.Ra. and A. K.; F.B. fabricated the microfluidic flow cell. N.R. performed sSMLM measurements. N.R. performed confocal microscopy measurements with the assistance of E.G.; N.R. analyzed the data and interpreted it with the help of J.C., E.G., M.L., B.Re., S.R. and A.R. ; K.W. and T.T. contributed materials; N.R. wrote the paper with the assistance of J.C. and inputs from all authors; A.R. supervised the project; All authors discussed the results and commented on the manuscript.

\section*{Competing Interests Statement} 
The authors declare no competing interests.

\renewcommand\refname{References}
\bibliography{references}

\section*{Methods}

\subsection*{Sample preparation}
Pristine hBN flakes from high-quality crystals\cite{taniguchi2007synthesis} were exfoliated onto borosilicate glass coverslips (no. 1.5 Micro Coverglass, Electron Microscopy Sciences, 25 mm in diameter, {170 µm thick), using low-adhesion blue tape. The glass coverslips were cleaned either by (i) sonication in acetone followed by rinsing in isopropanol (ii) sonication in 2\% Hellmanex III glassware cleaning agent solution, following by sonication in deionized water. No differences were observed between (i) and (ii). The coverslips were rinsed three times in the last solution, after which they were dried with a nitrogen gun covered with a 20 nm-pore filter. Adhesion to the substrate upon exfoliation was promoted by oxygen plasma cleaning of the coverslip (2 min, 100W).} 
High-quality crystals purchased from HQ Graphene {and hBN crystals obtained by crystallization of hBN from molten iron in nitrogen-hydrogen atmosphere} were also tested, and exhibited no notable difference with samples grown by the authors. The crystals were immersed immediately following exfoliation, and the PEEK chamber was thoroughly rinsed three times at each solvent exchange with fresh solvent. { The chamber was covered by an additional glass coverslip to prevent solvent evaporation and contamination.} The fabrication details for the microfluidic flow cell are given in the corresponding Supplementary Methods. Solvents used are detailed in Supplementary Table 1. 

Nanoslits are made of a van der Waals (vdW) heterostructure of a spacer layer sandwiched between a top layer and a bottom layer following the same protocol as previously reported \cite{radha2016molecular}. Here, the vdW stack is composed of top mica-graphene spacer-bottom hBN. In brief, thin (few atomic layers) graphene was first patterned via EBL into parallel strips with a width of 1 µm and a separation of 150 nm. A mica crystal ($\approx$ 200 nm thick) was then transferred on top of the graphene spacer, via PMMA based transfer method. Then this mica-graphene spacer stack was lifted and transferred onto a freshly exfoliated hBN layer. This whole stack was then transferred onto a glass coverslip for the imaging. The slit dimensions of the final device is shown in Figure 5a where it has a width of approximately 150 nm, a length of 20 µm and a height equivalent to the thickness of the graphene spacer. Fabrication flow chart and materials are described in detail in Supplementary Figure 15.

\subsection*{Chemicals used}
All chemicals were purchased with maximum purity grade available, and some were purchased anhydrous. The full list is provided in Supplementary Table 1. No effect of solvent purity or residual water traces on hBN fluorescence activation was observed.

\subsection*{Optical microscopy}
Widefield imaging was performed on a custom wide-field fluorescence microscope, described elsewhere\cite{comtet2019wide}. Briefly, the emitters are excited using 561 nm laser (Monolithic Laser Combiner 400B, Agilent Technologies), which is collimated and focused on the back focal plane of a high-numerical aperture oil-immersion microscope objective (Olympus TIRFM 100X, NA: 1.45). This configuration leads to wide-field illumination of the sample in a circle with  $\approx$ 25µm diameter. Fluorescence emission from the sample is collected by the same objective and spectrally separated from the excitation light using dichroic and emission filters (ZT488/561rpc-UF1 and ZET488/561m, Chroma) before being projected on an EMCCD camera (Andor iXon Ultra 897) with EM gain of 150. An additional spectral path, mounted in parallel to the localization path allows for simultaneous measurements of the emission spectra from individual emitters (see details below). The sample itself is mounted in a sealed fluidic chamber, which is placed on a piezoelectric scanner (Nano-Drive, MadCityLabs) for fine focus. Typical exposure time is 20-50 ms and typical laser power 10-100 mW for the widefield excitation area of  2$\times10^3$µm$^2 $, resulting in a power density of $0.35-3.5$ kW/cm$^2$. Unless mentioned otherwise, the illumination power density was set to 3.5 kW/cm$^2$. A typical acquired image stack contained 2-10 thousand frames.

Confocal microscopy measurements were performed in an inverted microscope configuration allowing to image samples in liquid in custom-made chambers made of a glass ring glued onto the glass coverslip using epoxy resin (Araldite). The excitation laser was a 561 nm picosecond diode laser (Sepia II, PicoQuant) used either in pulsed or CW mode. The emission, after collection with a water-immersion microscope objective (Nikon SR Plan Apo 60x, NA: 1.27) was split between two fiber-coupled APDs (SPCM-AQRH, Excelitas) in a Hanbury Brown and Twiss configuration. Photon correlation measurements were performed using the PicoHarp TCSPC module (PicoQuant).

\subsection*{sSMLM procedure}

Acquired image stacks from the wide-field microscope were processed using ThunderSTORM\textsuperscript{41}. 
The localization uncertainty of emitters is given by $\sigma_\text{loc} \sim \sigma_\text{PSF}/\sqrt{N_\text{loc}}$, where $\sigma_\text{PSF}$ represents the radial extent of the microscope point spread function (approximately 110 nm), and $N_\text{loc}$ is the number of photons counted in the diffraction-limited spot. All symbols used in the manuscript are defined in Supplementary Table 2. In our imaging conditions, localization uncertainties ranged from 10 to 20 nm. 
Localizations used for emitter counting, super-resolution image rendering and trajectory analysis were filtered to keep only localization events with $30<\sigma_\text{PSF}<200 \text{ nm}$ and filter out artifacts. To generate super-resolved images, individual SMLM localization events were rendered as 2D Gaussians with a standard deviation of 20 nm unless specified otherwise, reflecting the typical uncertainty on the emitter’s position.

{Spectral SMLM (sSMLM) was performed following a procedure described in our previous work\cite{comtet2019wide}, which we summarize here. Emitters in the spatial channel are localized using the ImageJ plugin ThunderSTORM. Briefly, a wavelet filter is applied to each frame, and peaks are then fitted by 2D-integrated Gaussians. The obtained localizations $(x_\text{loc}, y_\text{loc})$ are matched to a reference spectral point (wavelength of 650 nm) through a matrix transformation of the form $(x_\text{spec}, y_\text{spec})=A\times(x_\text{loc}, y_\text{loc})+B$. $A$ is a 2×2 matrix; and $B$ is a vector. The spectrum is then extracted as the vertical profile around $(x_\text{spec}, y_\text{spec})$ in a box of 6x40 pixels centered at the emitter. Pixel values are translated as spectra intensities through the linear relationship $\Delta y_\text{spec} \approx a\times \lambda$, with $a\approx$0.25px/nm. This value and transformation matrix coefficients were obtained using fiducial markers emitting at known wavelengths. As described in our previous work, this was achieved by using broadband emitters (fluorescent beads) and narrow bandpass filters in the emission path. The signal-to-noise ratio of the single-emitter spectrum is evaluated and all spectra passing a threshold are averaged to yield and ensemble spectrum. We typically used 3000 frames for each solvent and set the threshold to $I_\text{loc}>300$ photons in all measurements.}

Ensemble sSMLM spectra were maximum-normalized and fit using Python package LMFIT\textsuperscript{42}, to a model composed of 2 Lorentzians corresponding to the ZPL and PSB and a linear background. As shown in Fig. 5g-h, only the part of the normalized signal emerging more than 20\% over the background was fit, as the background-induced spectra tails could pose fitting issues. For spectra in nanoslits, a spatial filter was applied to avoid counting signal from contamination in between nanoslits (details in Supplementary Fig. 18). Error bars on ZPL and PSB wavelength (Fig. 3c) as well as ZPL-PSB detuning (Supplementary Fig. 12) were obtained by evaluating the standard deviation of the parameters obtained from fitting the average of randomly chosen groups of 100 single-emitter sSMLM spectra. This ensured sufficient signal-to-noise ratio for fitting while reflecting variations within the ensemble spectra.

\subsection*{Trajectory analysis}
     Wide-field fluorescence frames acquired with the EMCCD camera were first localized using ThunderSTORM\textsuperscript{41}. Trajectories were obtained by applying the Crocker-Grier linking algorithm\textsuperscript{43} to the localization microscopy tables. We used the implementation of the algorithm provided by the Python package trackpy\textsuperscript{44}. Briefly, for each localization event at a given frame, the algorithm links another event if it is found at the next frame within a specified search range. This search range was set to 120 nm in Fig. 2, as the probability of having a 1D displacement exceeding this value is less than 1\% according to the analysis in Fig. 2e.

    In the statistical analysis of trajectories, the emitter position is described as a random variable of time $X(t)$. The trajectory residence time on the crystal surface $T_\text{res}^T$ corresponds to the duration of this single trajectory. The one-dimensional displacement probability distribution function is given by PDF$(x,\tau)=P\Big(X(t+\tau)-X(t)=x\Big)$. The residence times of molecules at single defect sites $T_\text{res}^D$ were obtained using the same linking algorithm with a linking range of 35 nm (about twice the typical localization uncertainty) instead of 120 nm.

\section*{Data availability}
The data used to produce the graphs is provided as Source Data Files 1-5 for the main text figures, and Supplementary Data 1 for the Supplementary Figures. The linked localization table corresponding to Supplementary Video 1 and Figure 2 and the corresponding raw frames are available on Zenodo at \href{https://doi.org/10.5281/zenodo.8087398}{https://doi.org/10.5281/zenodo.8087398}.

\section*{Methods-only references}
\textbf{41}. Ovesn\`{y}, M., K\v{r}ížek, P., Borkovec, J., Švindrych, Z. \& Hagen, G. M. ThunderSTORM: a comprehensive ImageJ plug-in for PALM and STORM data analysis and super-resolution imaging. \textit{Bioinformatics} \textbf{30}, 2389–2390 (2014).\\
\textbf{42}. Newville, M., Stensitzki, T., Allen, D. B. \& Ingargiola, A. LMFIT: Non-Linear Least-Square Minimization and Curve-Fitting for Python, DOI: 10.5281/zenodo.11813 (2014).\\
\textbf{43}. Crocker, J. C. \& Grier, D. G. Methods of digital video microscopy for colloidal studies. \textit{J. Colloid Interface Sci.} \textbf{179}, 298–310 (1996)\\
\textbf{44}. Allan, D. B., Caswell, T., Keim, N. C., van der Wel, C. M. \& Verweij, R. W. soft-matter/trackpy: v0.6.1, DOI:
10.5281/zenodo.7670439 (2023).

\flushbottom

\thispagestyle{empty}

\newpage




{\LARGE \textbf{Supplementary Information}}\\

\tableofcontents 

\renewcommand{\figurename}{Supplementary Figure}
\setcounter{figure}{0}    
\section*{Supplementary Videos Legend}
\addcontentsline{toc}{section}{Supplementary Videos Legends}

\textbf{Supplementary Video 1} \\
Wide-field movie of the crystal presented in Figure 1 immersed in ethanol, corresponding to the data presented in Figure 1b-d. The crystal is initially in the dark, and the continuous 3.5  kW/cm$^2$ illumination is turned on at the beginning of the movie, inducing a decrease in the density of emitters as quantified in Figure 1d. The original images were acquired with 10 ms exposure time, but here we combined frames to present a lighter movie with a higher signal to noise ratio, at the expense of a slower sampling rate (50 ms).  Scale bar: 2 microns.\\

\noindent
\textbf{Supplementary Video 2}\\
Representative wide-field movie of another hBN crystal in isopropanol, taken from the steady state under 3.5 kW/cm$^2$ illumination. This data set was used for Figure 2. The exposure time was initially set to 6 ms but frames were combined to obtain a similar sampling rate as above and a lighter movie. Scale bar: 2 microns. The full raw data is available on Zenodo at \href{https://doi.org/10.5281/zenodo.8087398}{https://doi.org/10.5281/zenodo.8087398}. \\

\section*{Supplementary Methods}
\addcontentsline{toc}{section}{Supplementary Methods}

\subsection*{Macroscopic flow response of liquid-activated hBN emitters}
\addcontentsline{toc}{subsection}{\hspace{1 cm}  Macroscopic flow response of liquid-activated hBN emitters}

We monitored the fluorescence of hBN crystals while driving liquid flow over their surface. For this purpose, we fabricated a microfluidic flow cell onto a coverslip, allowing to perform wide-field fluorescence imaging of a crystal encapsulated in a microchannel in which  ethanol was flushed using a syringe pump (PHD Ultra, Harvard Apparatus). The flow cell was custom designed using Secure-Seal silicon imaging spacers (Grace Bio-Labs) with a thickness of H=120 µm, in which a $\approx$1 mm opening was made to define the channel. The spacers were sandwiched between two coverslips, and inlet and outlet tubings were connected using epoxy resin (Araldite). A sketch of the flow cell is presented in Supplementary Figure \ref{fig:figSIflow}a. The top coverslip was omitted for clarity. By setting flow rates in the range of tens to hundreds of microliters per minute, we were able to induce laminar flows in the range of centimeters per second over the crystal, as illustrated in Supplementary Figure \ref{fig:figSIflow}b. The surface flow rate, defined by molecular slippage, is considerably reduced by a factor of approximately $\ell_S/H \approx 10^{-4}$ where $\ell_S$ is the slip length of ethanol on hBN, which can be expected to be just a few nanometers$^{1,2}$. As shown in Supplementary Figure \ref{fig:figSIflow}c-d, the number of emitters was found to undergo a considerable increase when the crystal was submitted to  100 µL/min flow rate. Both super-resolved images were obtained by processing 1000 frames, under 1.5 kW/cm$^2$ illumination. The frame-wise localization microscopy counting of emitters on the crystal under flow conditions is presented in the bottom panel of Supplementary Figure \ref{fig:figSIflow}e. The top panel shows the syringe pump flow protocol. We observed that, after a steep increase when submitted to flow, the crystal fluorescence returns to its steady state level when the liquid goes back to rest.

\subsection*{Nanoslit fabrication}
\addcontentsline{toc}{subsection}{\hspace{1 cm} Nanoslit fabrication}
Nanoslit devices are van der Waals (vdW) heterostructures composed of 3 layered two-dimensional crystals. All the 2D materials here are obtained by mechanical exfoliation, using adhesive tapes. The middle graphene spacer crystal is sandwiched between the top mica crystal ($\approx$200 nm thick) and the bottom hBN crystal ($\approx$20 nm thick) via vdW assembly. First, using e-beam lithography (EBL) and oxygen plasma etching, a few-layer graphene crystal is patterned into parallel strips with a separation of 150 nm. We then transfer the mica crystal on top of the graphene spacer, using a PMMA wet transfer method, as shown in Supplementary Figure \ref{fig:flowchart} (step \textbf{a}). When choosing the top layer, it is desirable to use reasonably rectangular shaped mica crystals so as to open the slit entry on either side of the top (depicted in the optical image in \textbf{a’}). The top mica layer thickness is in the range of 150-250 nm; as a mica layer thicker than this would be less adhesive to the spacer layer, whereas a thinner layer could sag into the slits. Next, we transfer the stack of mica-graphene layer onto a freshly exfoliated bottom hBN crystal shown in Step \textbf{b}. A large crystal size of hBN is chosen as a bottom layer, so that it allows to compare emitters in the confined 2D slit (masked by spacer and top layer) with the one masked by graphene spacer only. The thickness range of the bottom hBN crystal is chosen to be between 15 nm to 25 nm. \textbf{b’} shows an optical image of the nanoslits where the bottom wall is the hBN surface, and the top slit wall is the mica surface. The slit height is determined precisely by the thickness of graphene spacer layer. In this work, we fabricated nanoslits with heights of $\approx$1.4 nm and $\approx$2.4 nm, corresponding to the four and seven graphene layers, respectively. From the AFM micrograph of the graphene spacer, the slit width can be seen as $\approx$150 nm and slit height $\approx$2.4 nm. In the final step (step \textbf{c}), we transfer the three-layer stack of top mica/graphene spacer/bottom hBN onto a glass coverslip via the PMMA transfer method, for subsequent imaging by sSMLM. The glass coverslips have pre-patterned markers of gold/chromium (thickness $\approx$50 nm) made by photolithography and physical deposition.

\section*{Supplementary Discussion}
\addcontentsline{toc}{section}{Supplementary Discussion}

\subsection*{Recovery of the fluorescence}
\addcontentsline{toc}{subsection}{\hspace{1 cm} Recovery of the fluorescence}
In order to verify that the gradual decrease in number of emitters under constant illumination reported in Figure 1d does not correspond to a degradation of the hBN crystals, we verified that the crystal fluorescence recovers when the illumination stops. For this, we recorded wide-field images and alternated 3.5 kW/cm$^2$ illumination periods of 20 s with variable dark times ranging from 20 s to 1 hour, and counted the fluorescence intensity per EMCCD pixel (Supplementary Fig. \ref{fig:recovery}a). The inset of Supplementary Figure \ref{fig:recovery}a shows the illumination protocol. We found that \textit{(i)} the steady state remains at a constant level and \textit{(ii)} the transient state becomes brighter with increasing dark time. 
As the steady state mean intensity per localization event in such conditions is 200 photons (Supplementary Fig. \ref{fig:recovery}b), assuming that the fluorescence comes from emitters which contribute 200 photons per frame each, we could estimate the number of active defects per pixel by dividing the crystal intensity by this single-defect reference  (Supplementary Fig. \ref{fig:recovery}c).
We observed that this increase in crystal fluorescence scales sub-linearly with time (power law with exponent $\approx$0.8, reaching complete recovery in about an hour), yielding a saturation around 200 active defects per square micron. This yields an upper bound for a typical inter-defect distance of $\approx$ 70 nm, which is consistent with native defect densities measured by scanning tunneling microscopy in similar pristine crystals$^{3}$. In Supplementary Figure \ref{fig:figS14nomoredefects}, we further verify that the steady-state density of emitters was stable during this crystal recovery in the dark (at illumination times >100s). Indeed, regardless of the dark time, upon exposure to 3.5 kW/cm$^2$ illumination, the crystal fluorescence reaches a similar steady state level within 10-100s. This data is consistent with the mechanism described below, where the total defect number does not change, and where the dark time dependency comes from the slow activation of emitters.




\subsection*{Proposed activation/photobleaching mechanism}
\addcontentsline{toc}{subsection}{\hspace{1 cm} Proposed activation/photobleaching mechanism}
We attribute the fluorescence recovery after photobleaching to a slow chemisorption of liquid molecules (M) onto defects (D), resulting in activated defects (MD). An activated defect in its ground state (MD) can absorb a photon with frequency $\nu_\text{exc}$ to reach the excited state MD*. It rapidly decays back to the ground state while emitting a photon with frequency $\nu_\text{em}$ and can perform this cycle millions of times as shown by photon count rates in Figure 4a. However, there is a probability that the emitter undergoes an intersystem crossing to the triplet state MD$^T$, which is long-lived$^{4}$ and leads to the emitter extinction upon breaking of the covalent bond between the defect and the molecule. These last two steps lead to photobleaching, and the defect returns to its optically inactive state (D), while releasing a charged molecule (M$^C$) to the solvent. This charged molecule undergoes surface diffusion and has a significant probability to bind to a neighboring defect. This phenomenology implies that the charged molecule M$^C$ has a much greater reactivity with a defect D than an uncharged molecule M.
 \begin{center}
\centering
\includegraphics{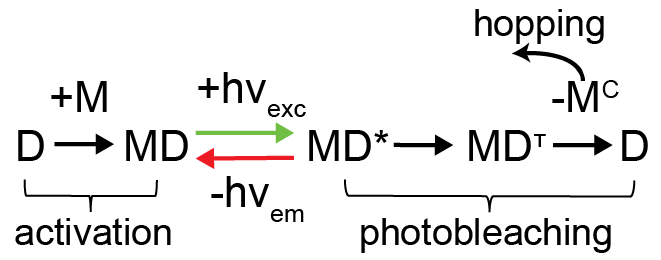}
\label{fig:figS13mechanism}
\end{center}

\subsection*{Absence of photoinduced damage}
\addcontentsline{toc}{subsection}{\hspace{1 cm} Absence of photoinduced damage}

We performed a characterization of crystals deliberately exposed to conditions harsher than those in this study (20 kW/cm$^2$ instead of 3.5 kW/cm$^2$). It seems likely that hypothetical light-induced damage, like the emission, should be located at the surface. We thus performed atomic force microscopy scans of a crystal which was partially exposed to intense laser light. The light intensity modulation was performed with a reflective spatial light modulator (HOLOEYE PLUTO-2) placed on the excitation path, to obtain a checkerboard pattern as shown below in Supplementary Figure \ref{fig:figS15afm}, where the image in (b) corresponds to localizations following the laser light exposure. The (white light) optical comparison of regions that were exposed vs. unexposed to laser light revealed no signs of light-induced damage (c). As imaged with atomic force microscopy, exposed and unexposed parts of the crystal were found to be identical in the height map (d,e), and more importantly, in the phase map of the crystal (f). AFM phase imaging did not reveal any difference between exposed and unexposed hBN, which is a qualitative indication of material uniformity$^{5,6}$. For comparison, we could resolve the ~3° phase change between glass and hBN (g). To check whether the overall crystallinity of the material could be affected in its bulk by the light, we verified through Raman spectroscopy of a thin (<10 nm) hBN crystal that the high-quality crystallinity was preserved even after 40 minutes of 20 kW/cm$^2$ illumination at 561 nm in ethanol. As shown in Supplementary Figure \ref{fig:figS16raman}, the full-width at half maximum (FWHM) of the B-N stretch mode was found to be 8.3±0.4 cm$^{-1}$. Raman spectroscopy was not able to distinguish exposed parts of the sample from pristine crystals, all of them showing the nominal FWHM in high-quality crystals, ~8 cm$^{-1}$, and far from strain and disorder-induced values (12 to 20 cm$^{-1}$)$^{7}$.

\subsection*{Vibrational analysis of the emitters}
\addcontentsline{toc}{subsection}{\hspace{1 cm} Vibrational analysis of the emitters}

{We extend here the analysis of the emission spectra to evidence further peculiar interactions between the excited defect and solvent molecules. As shown in Supplementary Figure \ref{fig:SIphononsideband}a, the energy difference between the ZPL and the PSB, which relates to the phonon dispersion of the material, was found to depend strongly on the solvent. The ZPL-PSB energy detuning ranges from 1255 cm$^{-1}$ for nonpolar pentane to  847 cm$^{-1}$  for polar methanol, further hinting that the phonon emission is affected by the dipolar nature of the liquid medium and, possibly, hydrogen bonding. Classically, the phonon dispersion should peak at vibrational modes of the hBN crystal, shown as dashed lines. Around 1365 cm$^{-1}$ is the most intense vibrational mode of hBN, B-N stretching, which is both Raman and IR-active, and around 820 cm$^{-1}$ is the IR-active out-of-plane B-N bending mode$^{8-10}$. The phonon side band of hBN defects PL spectra, which often arises in the 150-170 meV (1200-1400 cm$^{-1}$) detuning range, was attributed to B-N stretching previously$^{11,12}$. While some variations were observed throughout the literature, to the best of our knowledge, vibrational modes below 1200 cm$^{-1}$ were not found in purely solid-state defects.    Indeed, pristine hBN does not have any vibrational mode between 900 and 1200 cm$^{-1}$ which has been dubbed the 'phonon band gap' of hBN$^{13}$. Various bonds could yield a phonon energy in this range: functionalization of hBN nanosheets with oxygen gives rise to IR-active modes in this range$^{14}$. A recent theoretical study proposed that the chemisorption of carbon-bearing molecules on native hBN defects can give rise to vibrational modes precisely in the 900-1100 cm$^{-1}$ range$^{15}$. The formation of B-C bonds seems possible as they were found to occur around 1020 cm$^{-1}$ in boron-doped activated carbon$^{16}$ ,and carbon atoms are a common component of all activating solvents (type I and II).  The apparent lowering in phonon energy with increasing solvent polarity could be explained by the progressive redistribution of emitted phonons from B-N stretching to B-C stretching. This comes with an overall decrease of the phonon-assisted emission from 57\% down to 17\%, quantified in Supplementary Figure \ref{fig:SIphononsideband}b as the ratio of the integrated phonon side band over the integrated spectrum (inset). Overall, the vibrational signature of defects in organic solvents are compatible with covalent bonding between chemisorbed organic molecules and defect centers, as predicted by Lvova and Anina$^{17}$. This mechanism is similar to the activation of emitters through the covalent bonding of molecules to carbon nanotubes$^{18}$. However, as discussed in the main text, the residence time analysis at a defect suggests that this covalent bond is formed only transiently due to a decrease in desorption energy barrier by illumination.}

\newpage
\section*{Supplementary Figures}
\addcontentsline{toc}{section}{Supplementary Figures}

\addcontentsline{toc}{subsection}{\hspace{1 cm} Supplementary Figure 1. Spatial mapping of emitters in ethanol}
\begin{center}
\centering
\includegraphics[width=0.70\linewidth]{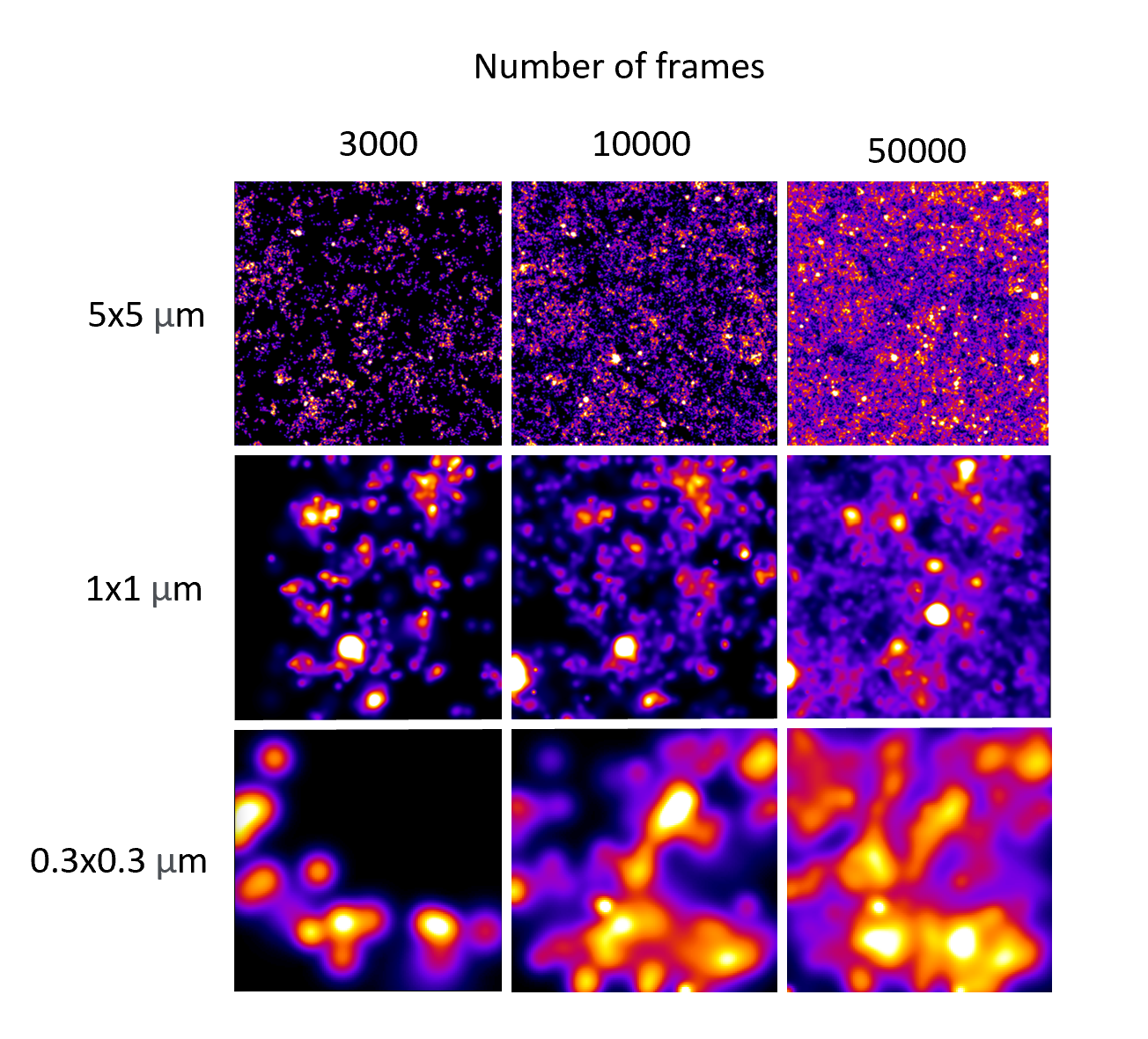}
\captionof{figure}{\textbf{Spatial mapping of emitters in ethanol}.  Super-resolved map of ethanol-activated emitters under 3.5 kW/cm$^2$ illumination for different imaging sizes ranging from 5$\times$5µm to 300$\times$300 nm and accumulation from 3k to 50k frames with 10 ms exposure. Each localization is rendered as a normalized Gaussian with a standard deviation of 15 nm to yield super-resolved probability maps used to visualize the spatial distribution of emission. We do not find a fully homogeneous defect distribution and heterogeneity is present at all scales, but there seems to be no consistent pattern appearing. As shown by the most magnified images (bottom), the dark region without localization events from the first 3000 frames eventually gets filled by localization events at later imaging times. We conclude that emitters are randomly distributed on the surface basal plane of hBN, and that the observed heterogeneities come from the long residence times at trapping sites, discussed in the main text.}
\label{fig:spatialMapping}
\end{center}

\addcontentsline{toc}{subsection}{\hspace{1 cm} Supplementary Figure 2: Characterizing the crystal surface integrity with atomic force microscopy}

\begin{center}
\centering
\includegraphics{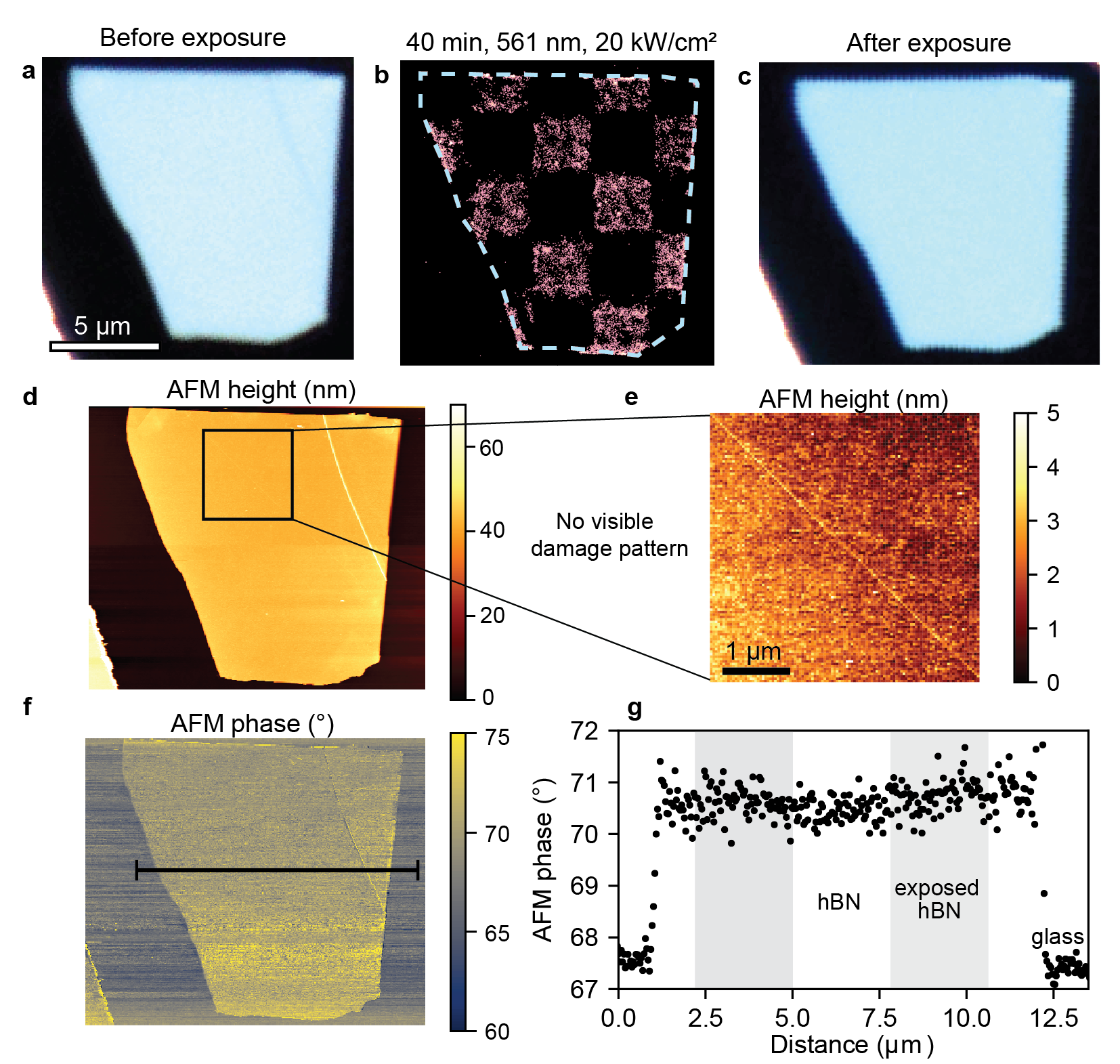}
\captionof{figure}{
\textbf{AFM characterization of the integrity of the crystal surface after exposure to light in liquid.} \textbf{a}, A pristine crystal was submitted to a spatially modulated 561 nm wide-field illumination of 20 kW/cm$^2$ for 40 minutes. Localization events from 1000 frames are presented as the super-resolved image in \textbf{b}, showing the checkerboard pattern of the spatial light modulator. No photoinduced damage was found by either optical microscopy \textbf{c} or atomic force microscopy height map \textbf{d,e} or phase map \textbf{f}. \textbf{g}, Phase profile along the line in \textbf{f}. The phase on the crystal is found to be uniform at around 70°, indicating no material change on the surface.}
\label{fig:figS15afm}
\end{center}

\addcontentsline{toc}{subsection}{\hspace{1 cm} Supplementary Figure 3: Characterizing the crystal bulk integrity with Raman spectroscopy}

\begin{center}
\centering
\includegraphics{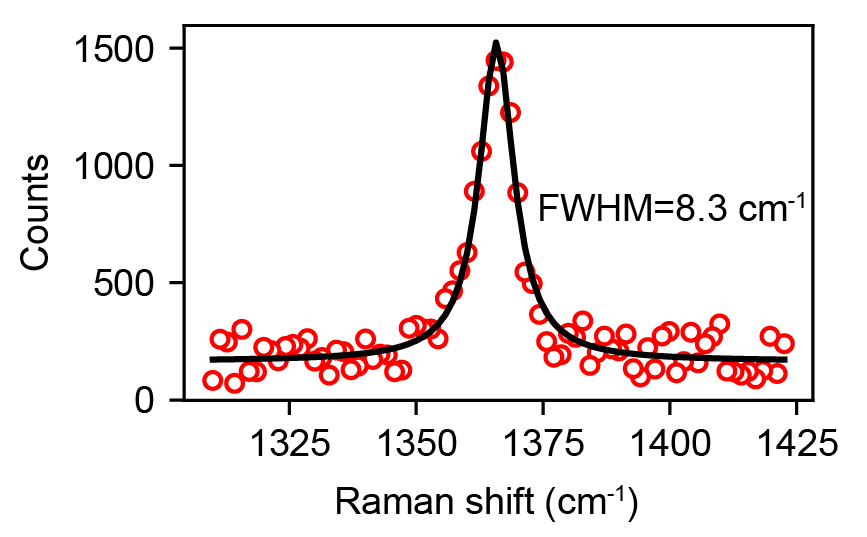}
\captionof{figure}{
\textbf{\textbf{Raman spectrum of hBN after intense exposure to light in ethanol. }} No difference is found compared with values reported for high-quality crystals in the literature$^{7}$. The solid line is a fit to a Lorentzian curve, used to obtain the full width at half maximum of 8.3±0.4 cm$^{-1}$.}
\label{fig:figS16raman}
\end{center}

\addcontentsline{toc}{subsection}{\hspace{1 cm} Supplementary Figure 4: Fluorescence recovery in the dark}

\begin{center}
\centering
\includegraphics[width=0.8\linewidth]{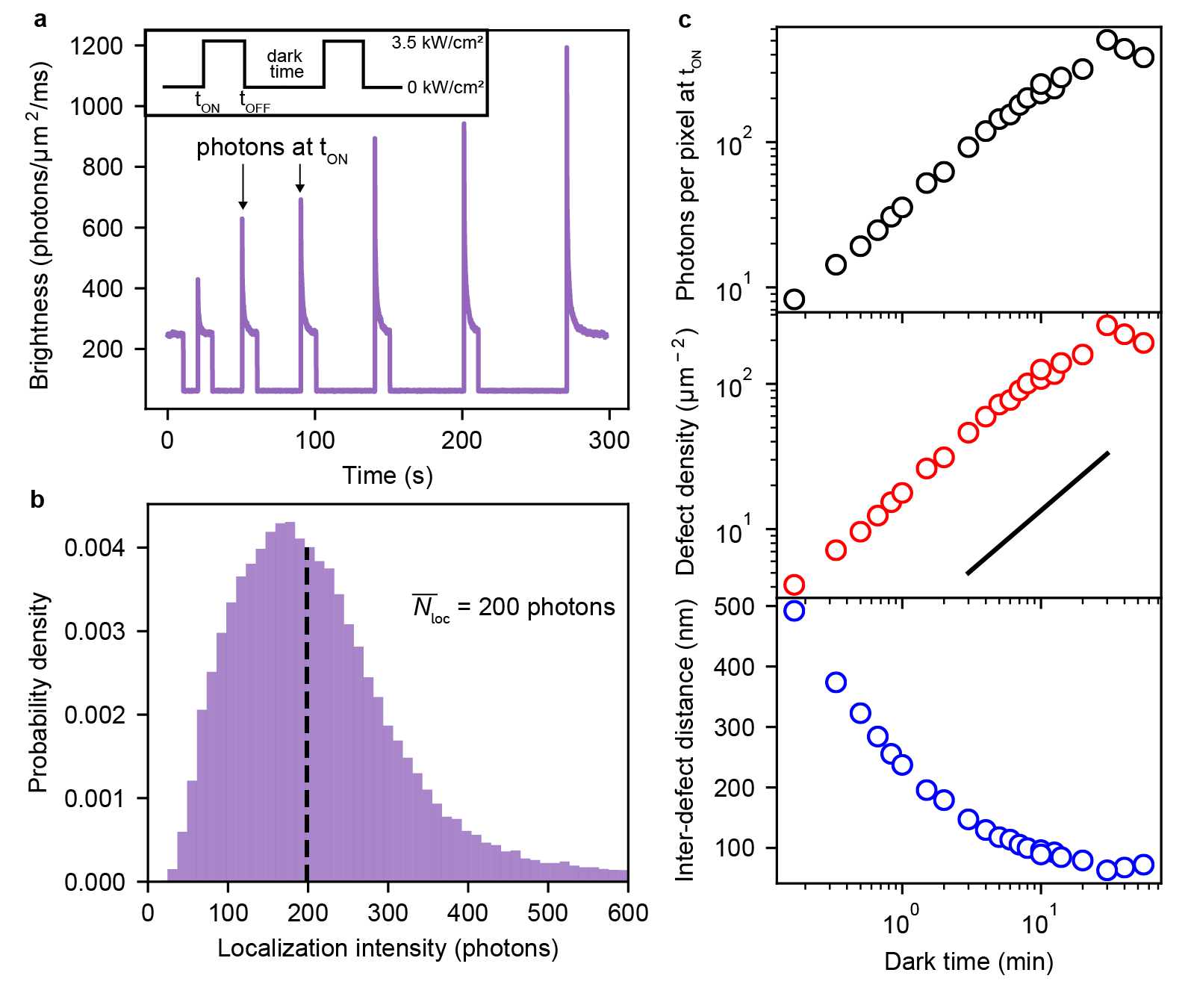}
\captionof{figure}{\textbf{Fluorescence recovery in the dark} \textbf{a}, Counting the fluorescent average camera signal per unit area as a function of dark time, while applying the illumination protocol defined in inset. The exposure time is 6 ms. \textbf{b}, Histogram of photon counts per localization in the steady state reached after 10 seconds of illumination, as shown in Figure 1d. \textbf{c}, Top: initial brightness, appearing as spikes in \textbf{a}, as a function of dark time. Middle: this brightness can be used to roughly estimate the defect density by dividing photon counts by the mean photon count indicated by a dashed line in \textbf{b}. Bottom: average inter-defect distance obtained from \textbf{b}, assuming randomly distributed emitters.
}
\label{fig:recovery}
\end{center}

\addcontentsline{toc}{subsection}{\hspace{1 cm} Supplementary Figure 5: Recovery in the dark: stability of the crystal fluorescence}
\begin{center}
\centering
\includegraphics{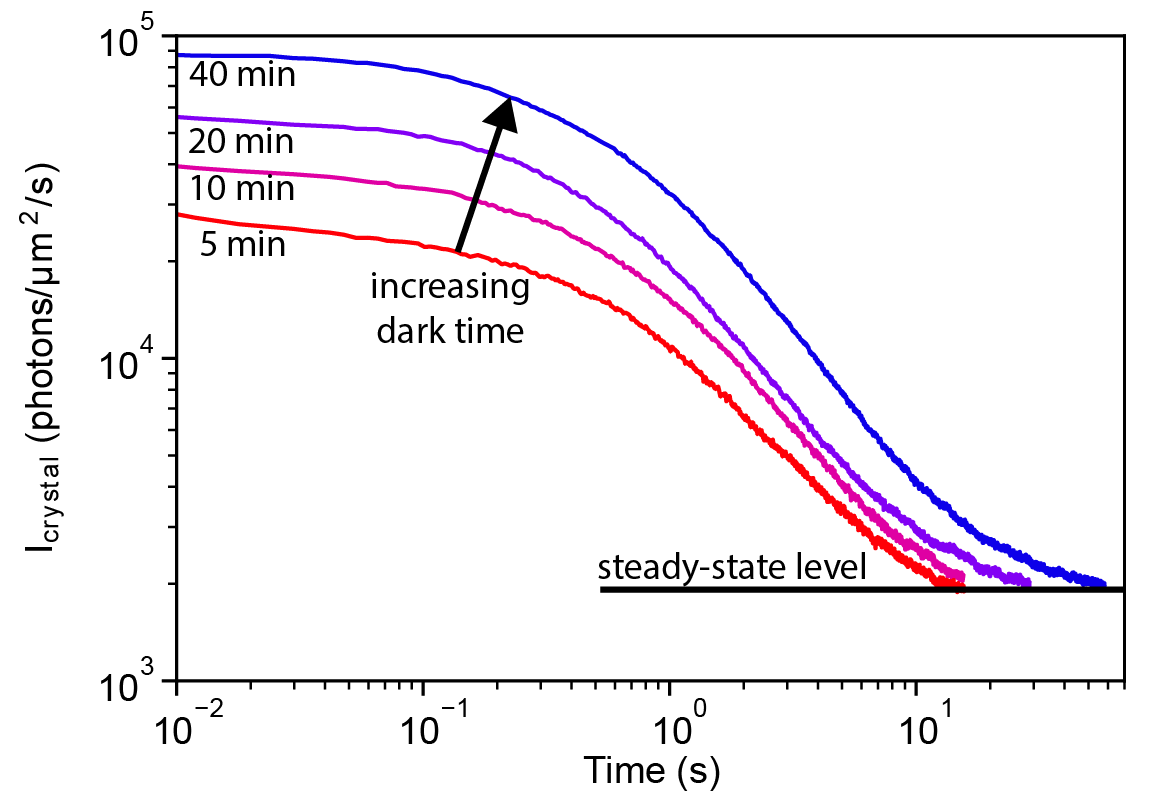}
\captionof{figure}{
\textbf{Recovery in the dark: stability of the crystal fluorescence. } Log-log plot of the bleaching curves corresponding to the data in Supplementary Figure \ref{fig:recovery}c, showing that the dark time for recovery increases the fluorescence in the early stage of the light exposure, but leaves the steady-state unchanged.}
\label{fig:figS14nomoredefects}
\end{center}

\addcontentsline{toc}{subsection}{\hspace{1 cm} Supplementary Figure 6: Emitter quenching by the addition of water to ethanol}

\begin{center}
\centering
\includegraphics{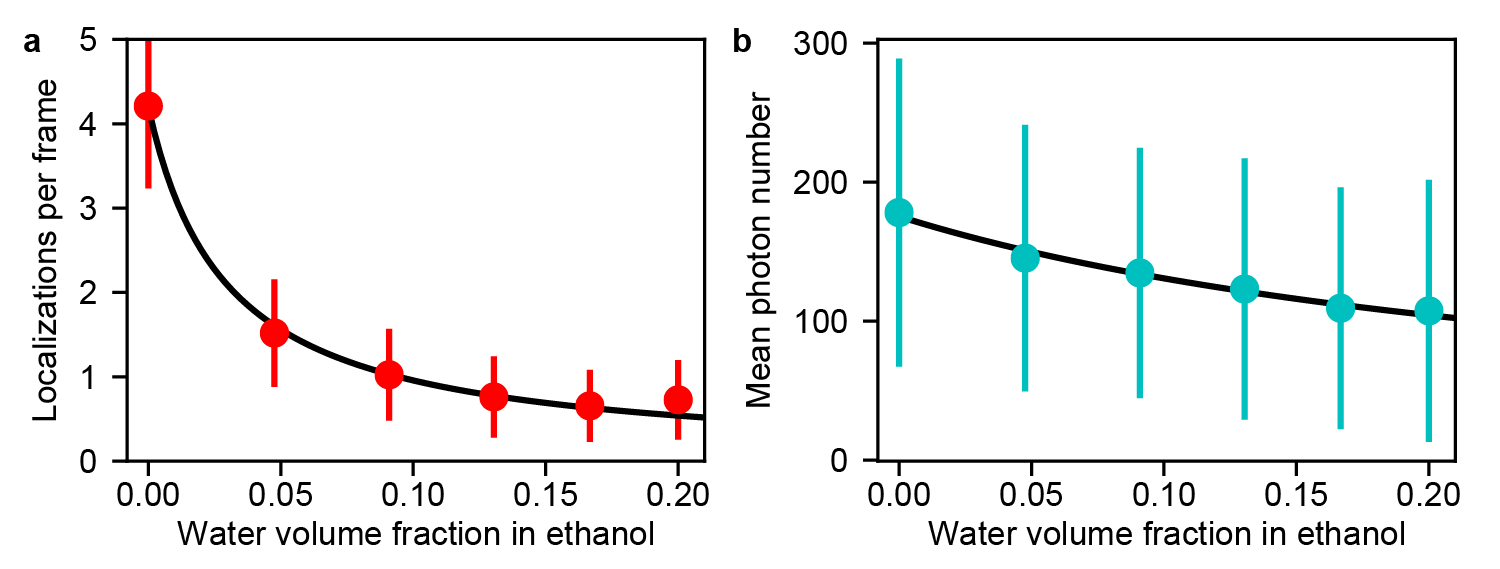}
\captionof{figure}{
\textbf{hBN fluorescence quenching by addition of water to the ethanol medium.} The overall fluorescence intensity drops as both \textbf{a} the localized emitter density and \textbf{b} the mean photon count of localizations drop. The experiment was conducted on a 13$\times$13 µm region of the crystal, illuminated at 1.5 kW/cm$^2$. The error bars correspond to the standard deviation of the distributions over the image (space) and frames (time).}
\label{fig:figS11quench}
\end{center}

\addcontentsline{toc}{subsection}{\hspace{1 cm} Supplementary Figure 7: Solvent comparison through SMLM emitter counting}
\begin{center}
\centering
\includegraphics[width=\linewidth]{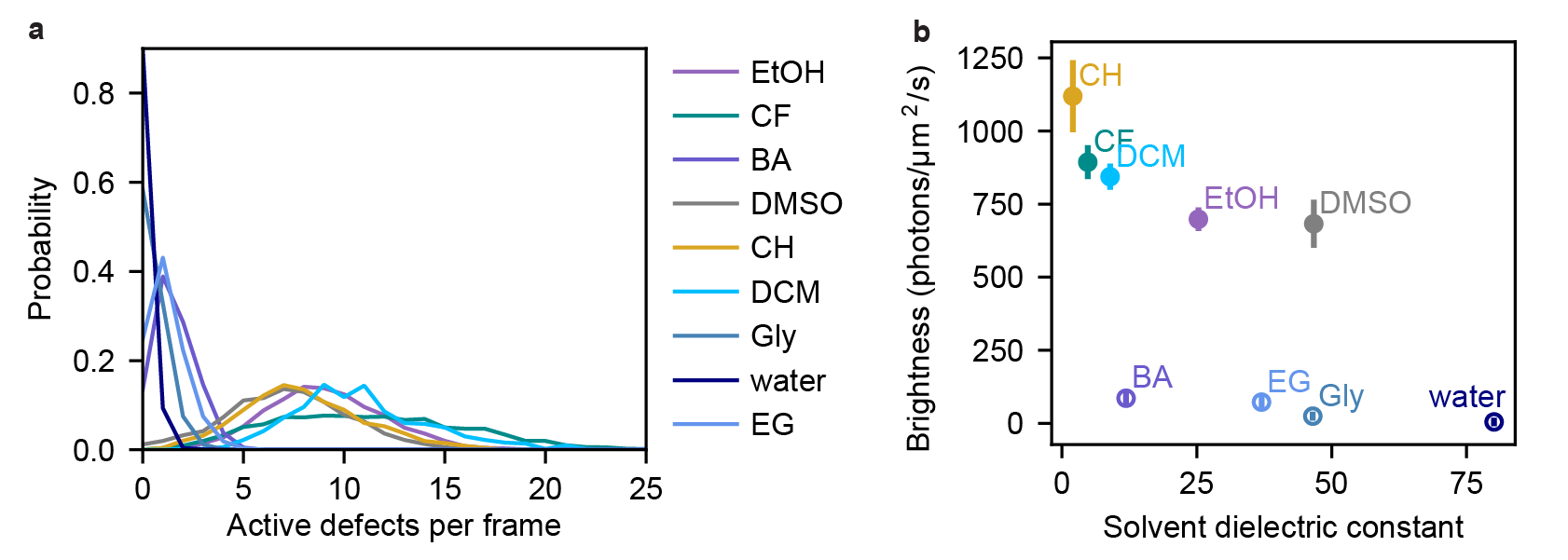}
\captionof{figure}{\textbf{Solvent comparison through SMLM emitter counting}. \textbf{a}, Probability densities of the number of active emitters per 13$\times$13 µm frame with 20 ms exposure. \textbf{b}, Crystal brightness $I_\text{crystal}$ as a function of the solvent dielectric constant, showing a clear separation between type I solvents (top) and type II and III solvents (bottom). The bar chart in Figure 1e presents a crystal intensity $I_\text{crystal}$ which comprises the localization intensity as well as the density of active emitters. Here the distributions of numbers of active emitters depending on the solvents are presented, clearly distinguishing type I solvents as having a high density of active emitters, unlike type II and III. Type II solvents have a non-zero average value which is better visualized in the logarithmic plot of Figure 1e. We further report a tuning of the crystal intensity with the solvent dielectric constant \textit{among type I solvents}, which correlates with spectral changes observed in Figure 3. However, the specificity of the phenomenon, distinguishing type I solvents from types II and III, cannot be explained in terms of dielectric constant only, and exhibits some chemical specificity. Error bars correspond to standard deviations over space and time.}
\label{fig:SolventExtraData}
\end{center}

\addcontentsline{toc}{subsection}{\hspace{1 cm} Supplementary Figure 8: Liquid activation in various organic solvents}

\begin{center}
\centering
\includegraphics[width=\linewidth]{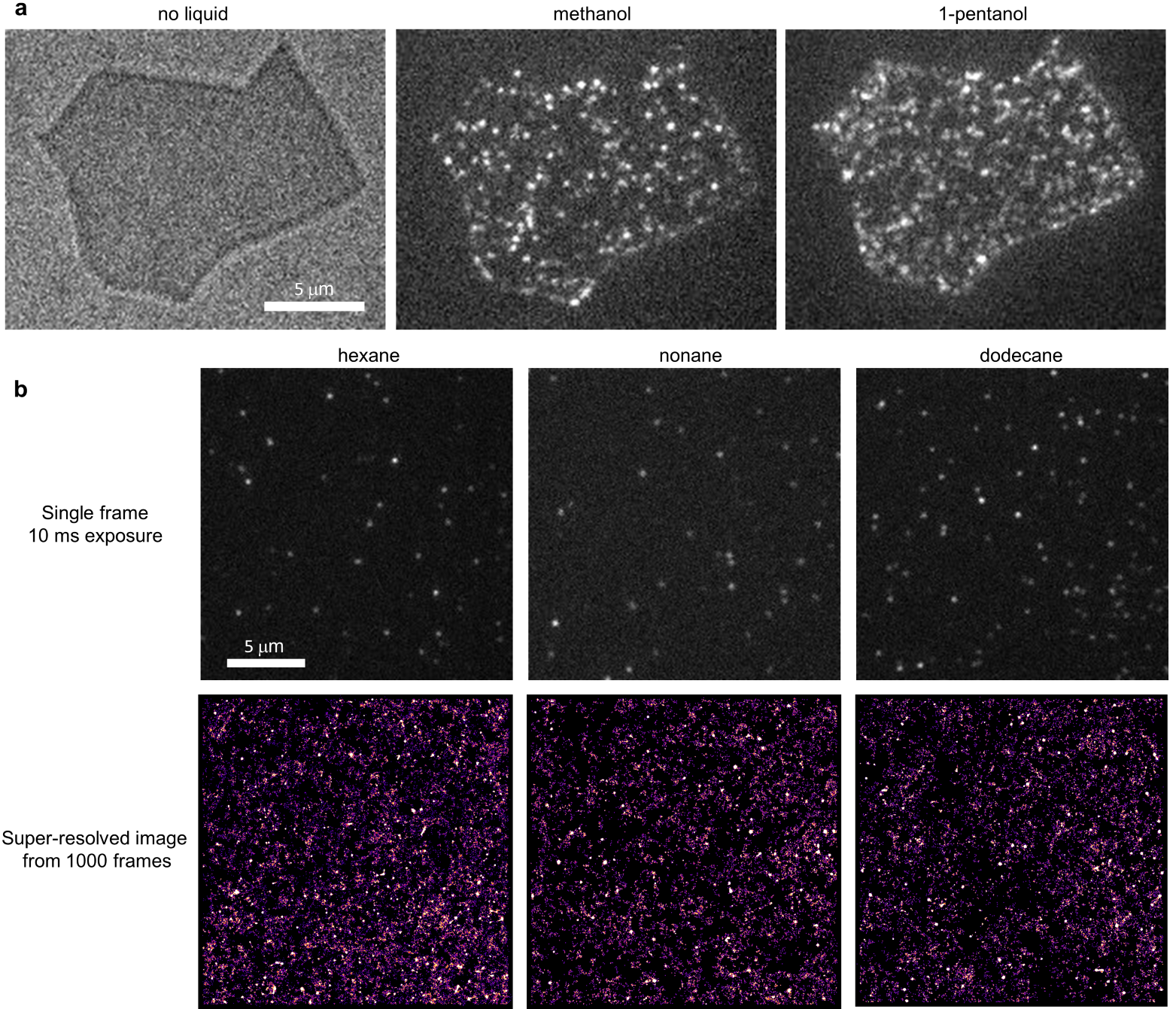}
\captionof{figure}{\textbf{Liquid activation in various organic solvents}. \textbf{a}, Images of the crystal from Figure 1 in air (white light illumination), in methanol and in 1-pentanol (561 nm laser illumination). \textbf{b} Emitters were counted from 1000, 10 ms exposure frames (top) and rendered as normalized Gaussians with a standard deviation of 15 nm to yield super-resolved probability maps used to visualize the spatial distribution of emission (bottom) for hexane, nonane and dodecane, respectively. The bright spots do not coincide between liquids and seem randomly distributed as was observed for ethanol in Supplementary Figure \ref{fig:spatialMapping}. No clear effect of the chain length is found. In Figure 1e solvents are used on freshly cleaved crystals the same day under the exact same illumination conditions. Here data are measured possibly on different days and with slight alignment-induced illumination changes.}
\label{fig:extraSolvents}
\end{center}

\addcontentsline{toc}{subsection}{\hspace{1 cm} Supplementary Figure 9: Solvent and light intensity dependency of the emitter density}

\begin{center}
\centering
\includegraphics[width=\linewidth]{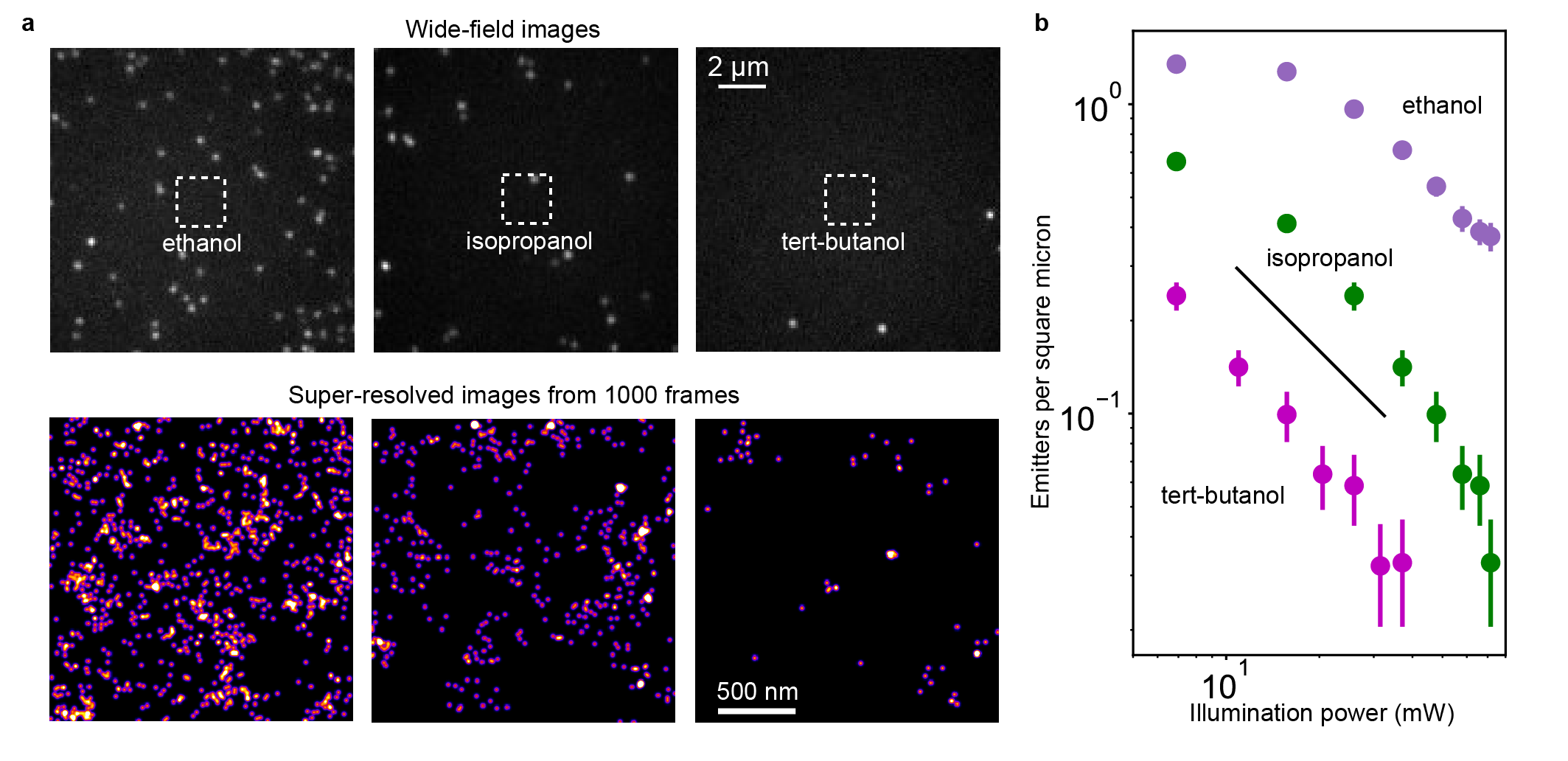}
\captionof{figure}{\textbf{Solvent and light intensity dependency of the emitter density}. We observed variations in emitter density depending on \textit{(i)} illumination and \textit{(ii)} the solvent used. We systematically varied both parameters for a subset of three solvents: ethanol (EtOH), isopropanol (IPA) and tert-butanol (TBA). \textbf{a}, Top: 30 ms exposure frames under 3 kW/cm$^2$ illumination in EtOH, IPA and TBA, respectively. Bottom: super-resolved images of the region enclosed by white dashes in \textbf{a}, reconstructed from 1000 frames and rendered as normalized Gaussians with size 15 nm. A decrease in emitter density is observed for increasingly substituted alcohols. This trend contrasts with the observations on n-alkanes and primary alcohols reported above.  \textbf{b}, Illumination power dependency of the emitter density for all three solvents presented on the left. The maximum power density obtained at the sample is about 3.5 kW/cm$^2$. The solid line indicates a slope of -1 corresponding to inverse proportionality. This light dependency is consistent with a lowered chemisorption energy at defect sites under illumination. In other words, the emitter deactivation should occur through the excited state as proposed in the mechanism in the Supplementary Discussion. Error bars correspond to the standard deviation over frames for the entire image.}
\label{fig:liquidsLight}
\end{center}

\addcontentsline{toc}{subsection}{\hspace{1 cm} Supplementary Figure 10: Macroscopic flow response of liquid-activated hBN fluorescence}
\begin{center}
\centering
\includegraphics[width=\linewidth]{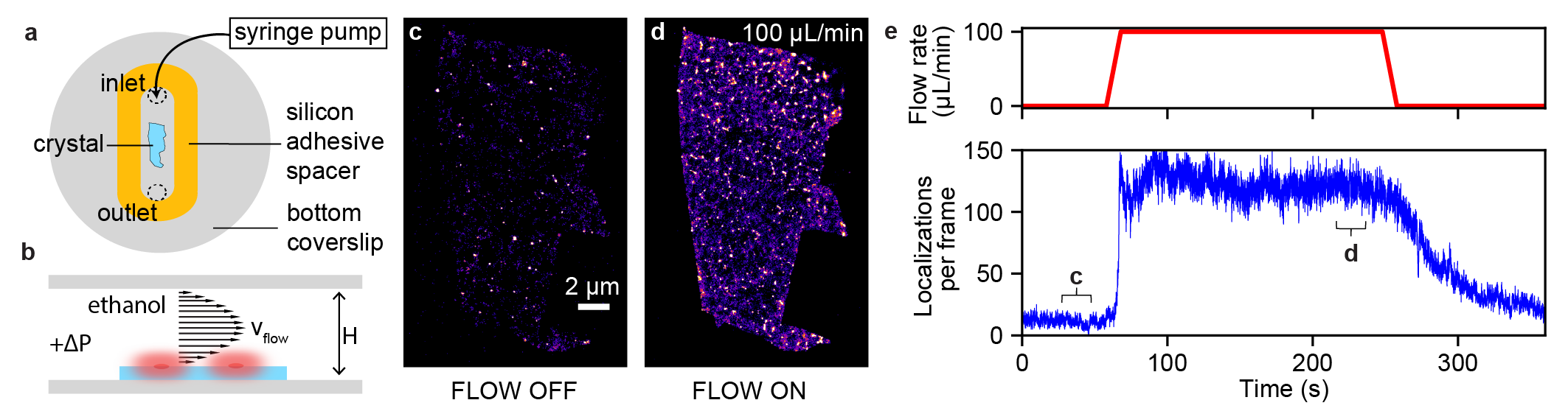}
\captionof{figure}{\textbf{Macroscopic flow response of liquid-activated hBN fluorescence}. \textbf{a}, Top view sketch of the flow cell. \textbf{b} Side view sketch of the flow profile in the microchannel. \textbf{c-d}, Localization microscopy images of a hBN crystal in ethanol (c) at rest and (d) under 100 µL/min flow rate, obtained from 1000 frames. \textbf{e}, Top: flow protocol programmed with the syringe pump. Bottom: localization microscopy counting of emitters while changing flow conditions. The regions used to render (c) and (d) are indicated on the bottom graph.}
\label{fig:figSIflow}
\end{center}

\addcontentsline{toc}{subsection}{\hspace{1 cm} Supplementary Figure 11: Long-lasting emitters \& polarization}
\begin{center}
\centering
\includegraphics[width=0.9\linewidth]{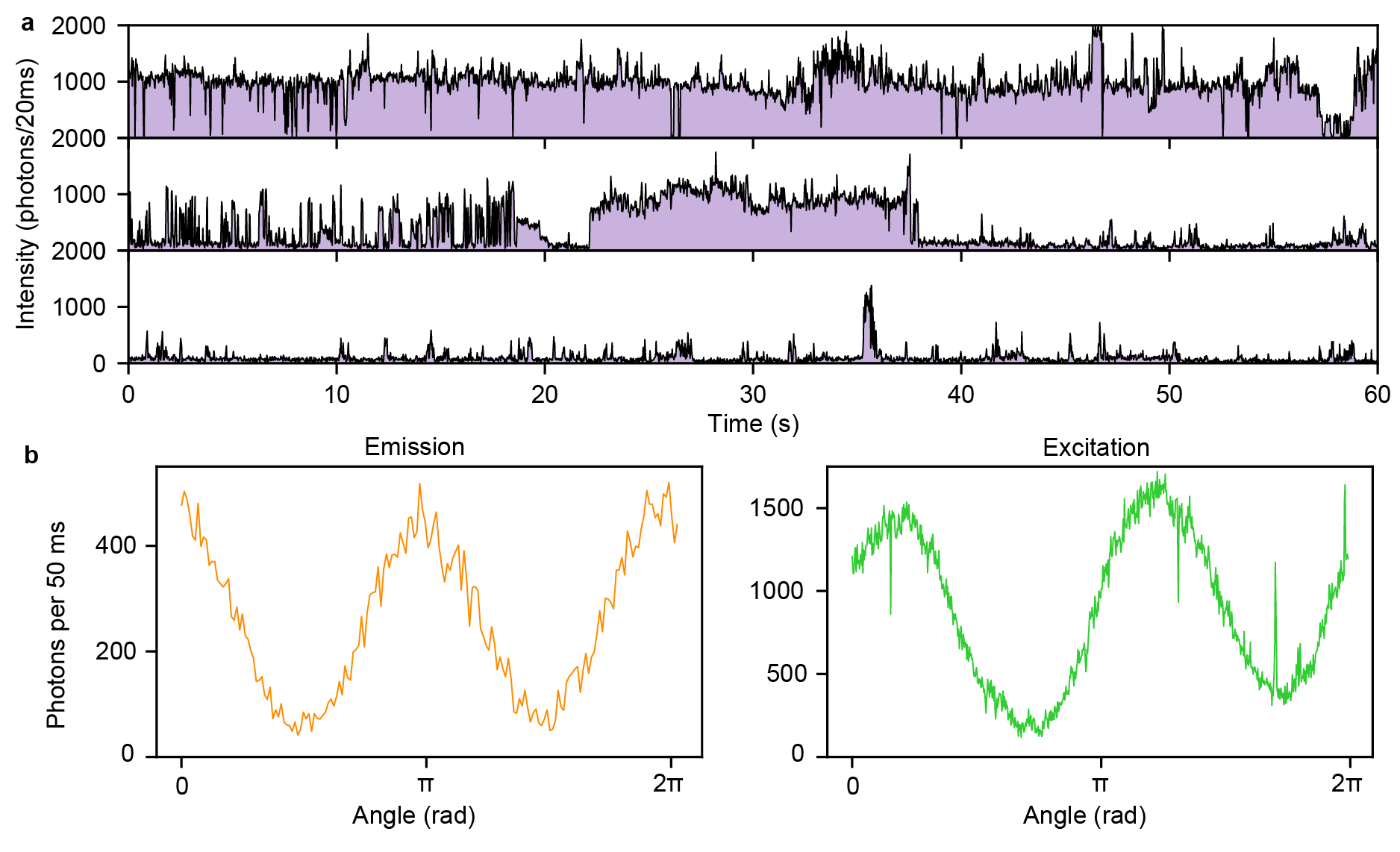}
\captionof{figure}{\textbf{Long-lasting emitters \& polarization}. \textbf{a}, Background-subtracted sum signal of 7x7 pixel boxes around ethanol-activated emitters under 3.5 kW/cm$^2$ illumination at 561 nm, showing different durations of emitter activation. While they appear as negligible in the residence times analysis as presented in Figure 2d, long-lasting traces exceeding 10 seconds were reliably found in type I solvents. \textbf{b} Using a long-lasting trace, the polarization response of emitters was measured by monitoring the fluorescent signal while rotating the polarization with a motorized half-wave plate placed before an analyzer (Thorlabs PRM1Z8, enabling rotation at 25 degrees per second). The excitation polarization response was analyzed in the same manner with the half-wave plate on the excitation path. The solvent used was pentane in both cases, and the illumination power were 0.35 and 1.4 kW/cm$^2$ for emission and excitation, respectively. The polar plot in Figure 3e was obtained by downsampling data to 50 angle values for clarity. Deviations from the perfect dipole fit in Figure 3e are likely due to the dichroic beamsplitter, whose birefringence was not compensated for.}
\label{fig:traces}
\end{center}

\addcontentsline{toc}{subsection}{\hspace{1 cm} Supplementary Figure 12: Vibrational analysis of the emitters}

\begin{center}
\centering
\includegraphics{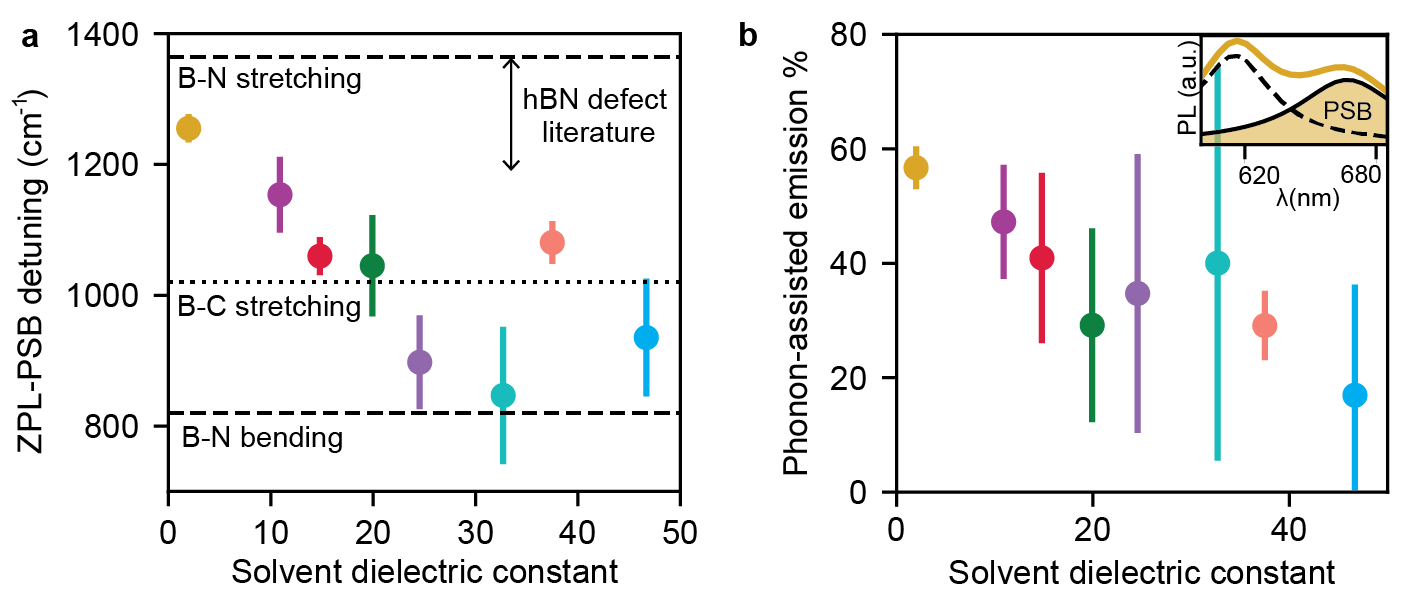}
\captionof{figure}{\textbf{Vibrational analysis of the emitters}. \textbf{a}, ZPL-PSB detuning as a function of the liquid dielectric constant, reporting on the phonon dispersion around emitters. Dashed lines correspond to the Raman modes of hBN present in this energy range while the dotted line corresponds to B-C bonds as possible candidates for the observed vibrations. \textbf{b}, Quantifying the relative magnitude of phonon-assisted emission with respect to direct emission as a function of the solvent dielectric constant. The ratio is estimated as the area of the integrated PSB divided by the full integrated spectrum, as shown in the inset. Error bars correspond to the standard deviation of fitting parameters obtained by fitting averages of 100 single-molecule spectra. }
\label{fig:SIphononsideband}
\end{center}

\vspace{2cm}

\addcontentsline{toc}{subsection}{\hspace{1 cm} Supplementary Figure 13: Affinity of emitters with dipolar molecules}

\begin{center}
\centering
\includegraphics{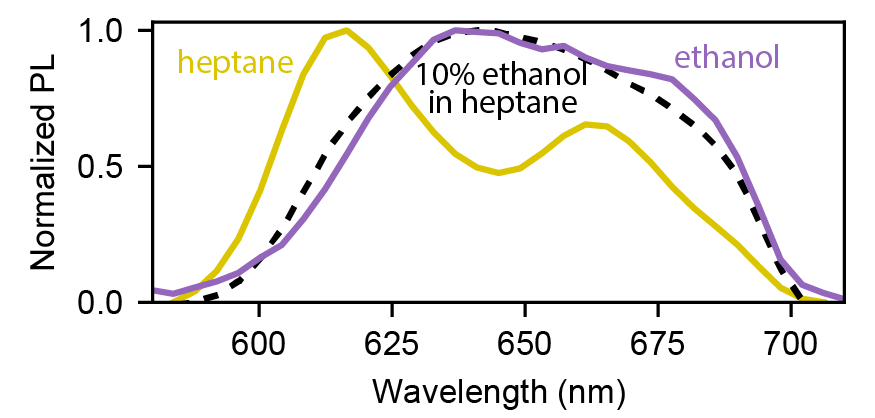}
\captionof{figure}{
\textbf{Affinity of emitters with dipolar molecules.} Comparison of the sSMLM spectra of heptane, ethanol and their 90:10 volume mixture. The resulting spectrum is very similar to that of ethanol, which is the minority species of the mixture. This result suggests that the charged defects possess a strong affinity towards polar solvents, leading to the enrichment in ethanol molecules within the emitter shell (diameter $\ell_\text{dip}\approx$1 nm). This phenomenon, known in the dye literature as preferential solvation$^{19}$, should allow sensing of polar molecules in apolar environments.}
\label{fig:figS12mix}
\end{center}

\addcontentsline{toc}{subsection}{\hspace{1 cm} Supplementary Figure 14: Liquid dependency of the Jablonski diagram}

\begin{center}
\centering
\includegraphics{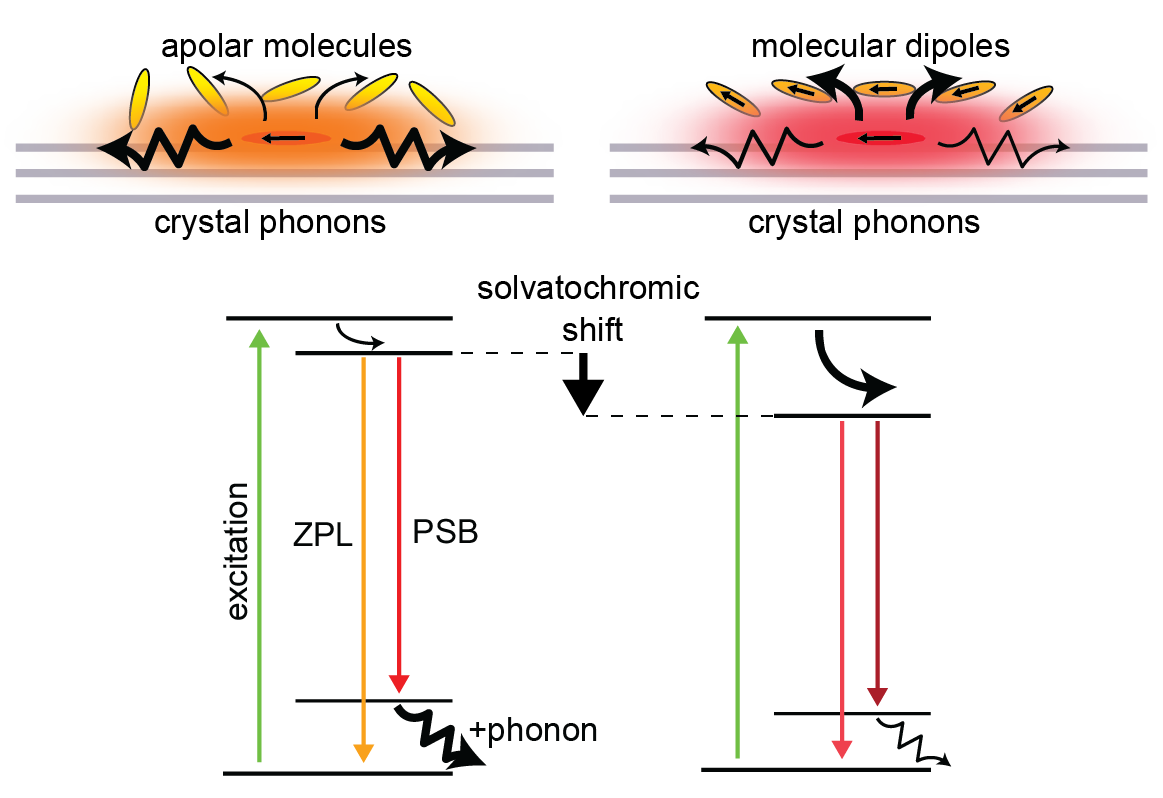}
\captionof{figure}{
\textbf{\textbf{Liquid dependency of the Jablonski diagram.}} Left: in an apolar environment such as a pure alkane, the emitter dipole is not stabilized through dipole-dipole interactions, leading to a ZPL around 615 nm (Fig. 3). Right: in a polar environment such as acetonitrile, the emitter dipole is stabilized through dipole-dipole interactions leading to a reduction in the ZPL energy, which is observed as a spectral redshift to 640 nm. As shown in Figure 3h, in apolar solvents the emitter is more strongly coupled to crystal phonons (bold zigzag arrow), whereas in polar environments the emitter is coupled preferentially to liquid molecules (bold curved arrow). This sketch is a possible explanation for positive \textit{solvatochromism} expected to occur when the magnitude of the excited state dipole is larger than the ground state one, which is the case for $\pi - \pi^*$ transitions$^{19}$. Solvatochromic shifts typically follow the static dielectric constant, which remains valid up to the relevant GHz regime for solvents used here$^{20}$.}
\label{fig:figS18newJab}
\end{center}

\addcontentsline{toc}{subsection}{\hspace{1 cm} Supplementary Figure 15: Nanoslit fabrication}
\begin{center}
\centering
\includegraphics[width=\linewidth]{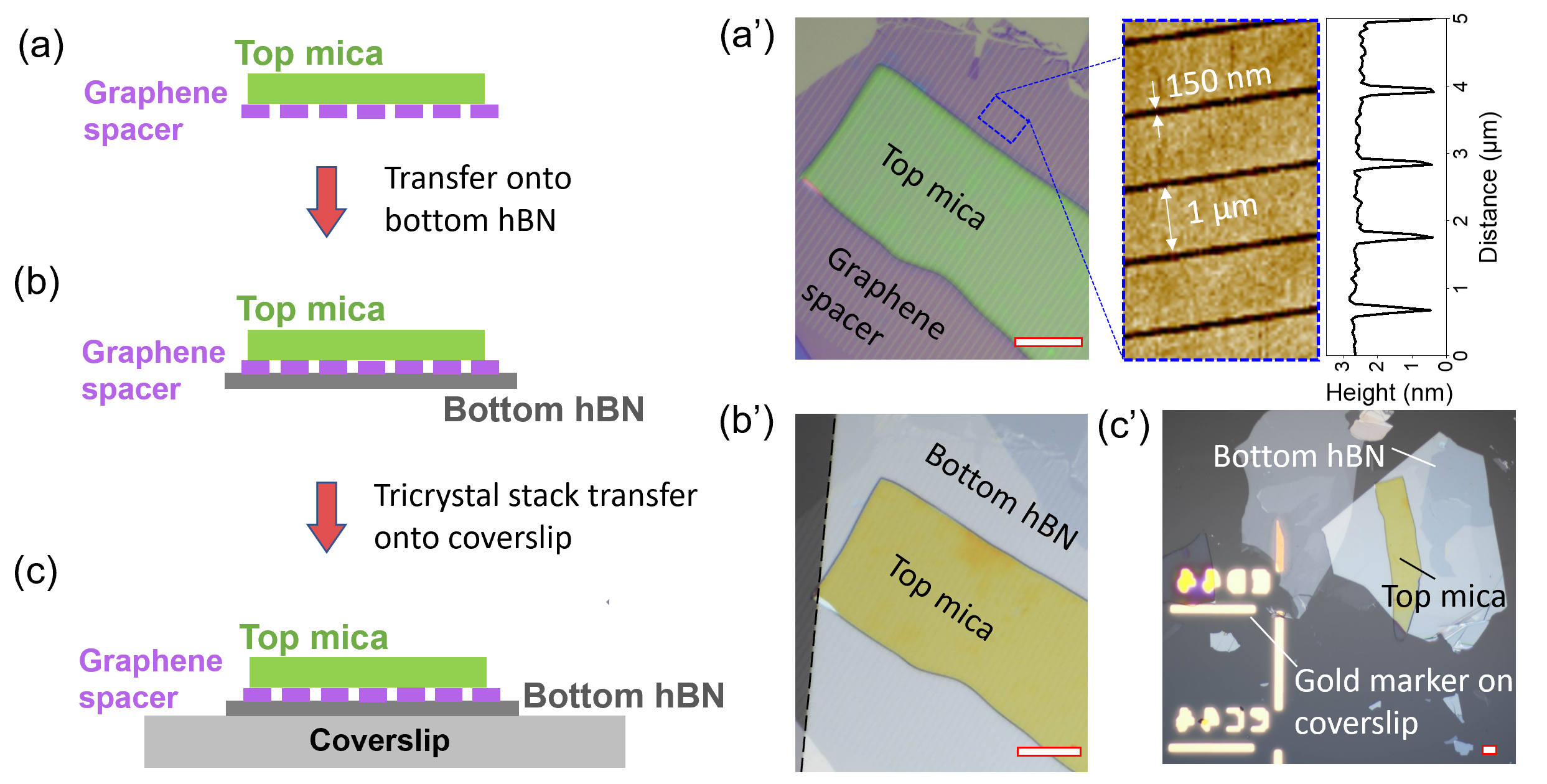}
\captionof{figure}{\textbf{Nanoslit fabrication}. Step \textbf{a}: Top mica layer transferred onto pre-patterned graphene spacer (by EBL). Step \textbf{b}: Transfer of mica-graphene stack onto a freshly exfoliated bottom hBN crystal. Step \textbf{c}: Transferring the heterostructure stack onto a coverslip with pre-defined gold markers. \textbf{a’}-\textbf{c’}: Optical images of mica-graphene stack, mica-graphene-hBN stack, and mica-graphene-hBN stack on coverslip, respectively. All scale bars, 10 µm. Next to \textbf{a’}, AFM micrograph of the graphene spacer from the area marked in the blue dashed rectangle is shown along with height profile.}
\label{fig:flowchart}
\end{center}

\addcontentsline{toc}{subsection}{\hspace{1 cm} Supplementary Figure 16: Comparison of bare, masked and slit-embedded hBN emitters}

\begin{center}
\centering
\includegraphics[width=0.6\linewidth]{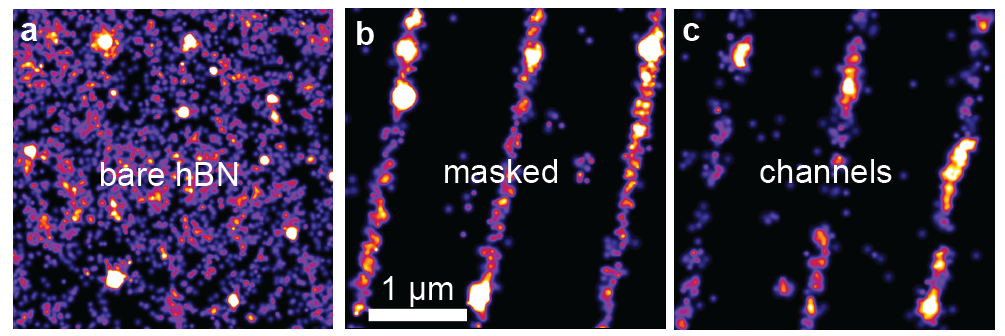}
\captionof{figure}{\textbf{Comparison of bare, masked and slit-embedded hBN emitters}. \textbf{a}, Super-resolved image from 10k frames of acetonitrile-activated emitters under 0.7 kW/cm$^2$ illumination with 20 ms exposure time. \textbf{b}, Graphene-masked hBN imaged in the same conditions. \textbf{c}, 2.4 nm-high nanoslit imaged in the same conditions. We evidence here that the uncovered part of masked hBN truly behaves as bare hBN, as shown by the agreement of bare and masked hBN spectra in Figure 5h-i. Nanoslits do exhibit a reduced number of emitters, but these emitters are not quenched and no photons are lost (less than 1\%), as shown by the histograms in Figure 5e. The heterogeneities observed here come from under-sampling of the bottom hBN surface defects due to the reduced number of active emitters, illustrated in Supplementary Figure \ref{fig:spatialMapping}. }
\label{fig:figSIchannels}
\end{center}

\addcontentsline{toc}{subsection}{\hspace{1 cm} Supplementary Figure 17: Effect of confinement on residence times}

\begin{center}
\centering
\includegraphics{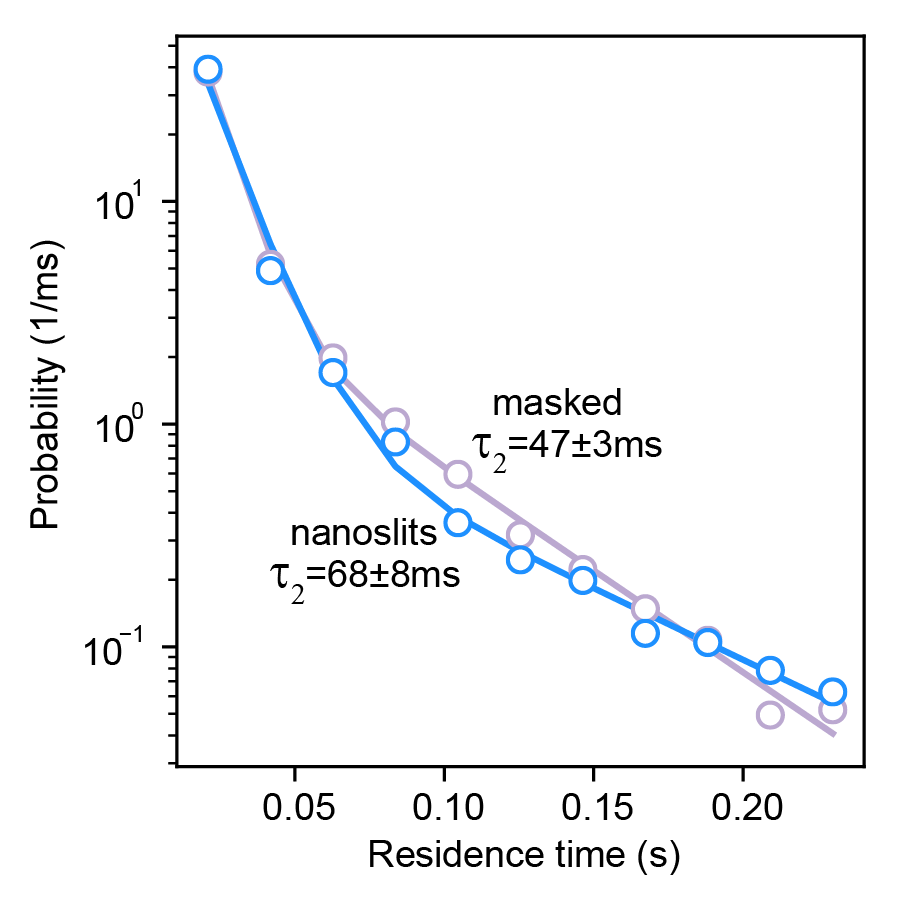}
\captionof{figure}{\textbf{Effect of confinement on residence times}.
Single-defect residence time $T_\text{res}^D$ distribution of emitters on masked hBN (purple) and nanoslits (blue) were fit to a two-component exponential decay as in Figure 3d. Residence times were obtained by applying the tracking procedure to the geometrically filtered emitters with an uncertainty-limited cutoff distance set to 35 nm for both datasets. Residence times were found to be affected by confinement as the nanoslit-confined emitters exhibited a long exponential decay time constant of $\tau_\text{res}^D=$68±8 ms while unconfined emitters on the same image had a $\tau_\text{res}^D=$47±3 ms time constant. This confinement-induced increase in residence time cannot explain the observed confinement-induced decrease in number of emitters shown in Figure 5e. This means that the leading effect of confinement is the decrease in activation rate, rather than the photobleaching kinetics. This observation echoes the result of the flow measurement, where the number of active emitters was found to increase dramatically when the fluid was driven across the crystal surface, suggesting that the exchange between bulk and surface liquid molecules is affecting the number of active emitters.}
\label{fig:figS17resconf}
\end{center}

\addcontentsline{toc}{subsection}{\hspace{1 cm} Supplementary Figure 18: Spatial filtering procedure for masked hBN and nanoslits}

\begin{center}
\centering
\includegraphics[width=8 cm]{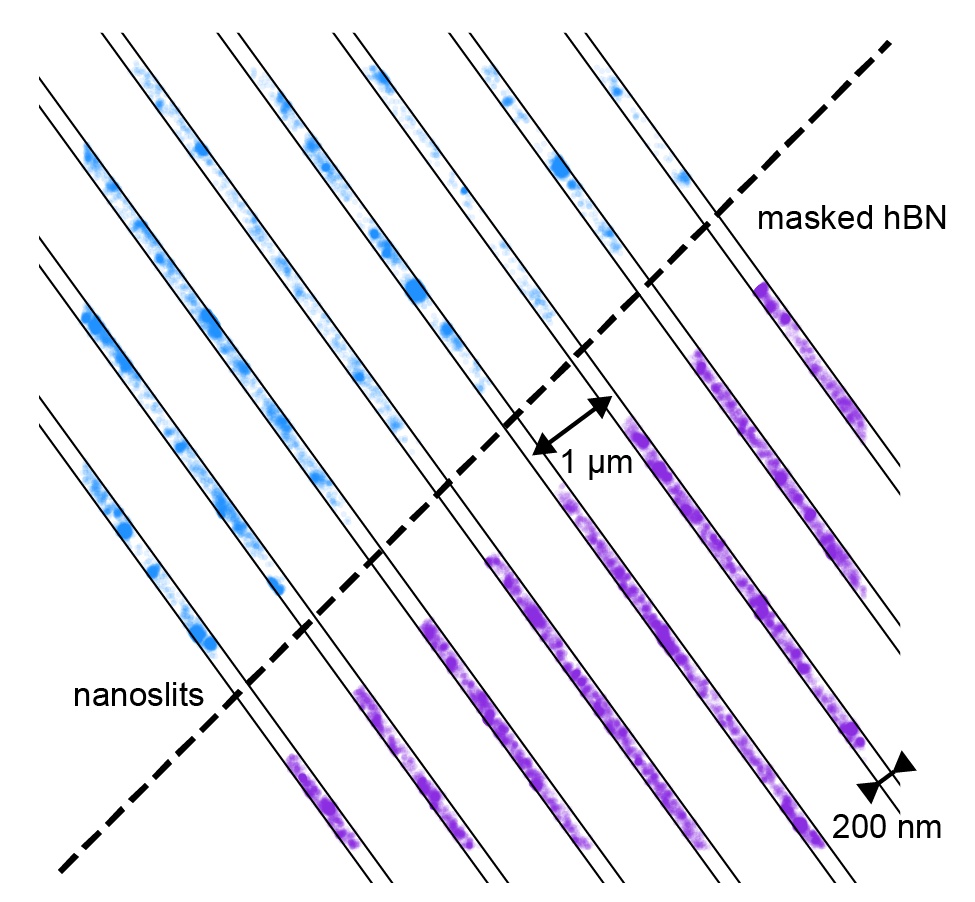}
\captionof{figure}{\textbf{Spatial filtering procedure for masked hBN and nanoslits}. Super-resolved image from the edge of the nanoslits used for the quantitative comparison performed in Figure 5e, overlaid with the mask applied for counting. The emitter density in nanoslits amounts to about one third of its value on bare hBN. For emitter counting and spectral assignments in samples with defined patterns, we filtered localizations spatially, effectively applying a post-processing mask matching the graphene pattern. The filter was set to 200 nm width to account for the slit width (150 nm) and twice the localization uncertainty of about 25 nm. This ensures that we did not assign counts or spectra to localization events occurring in between nanoslits, which can be seen in Supplementary Figure \ref{fig:figSIchannels}c for instance. These emitters are likely due to transfer-induced contamination, which is inevitable in fabrication but kept to low density. }
\label{fig:figSIchanmaskcrop}
\end{center}

\section*{Supplementary Tables}
\addcontentsline{toc}{section}{Supplementary Tables}
\subsection*{Supplementary Table 1: Chemicals used}
\addcontentsline{toc}{subsection}{Supplementary Table 1: Chemicals used}
\begin{tabularx}{0.8\textwidth} { 
  | >{\centering\arraybackslash}X 
  | >{\centering\arraybackslash}X 
  | >{\centering\arraybackslash}X | }
 \hline
 Solvent & Supplier & hBN activation type \\
 \hline
 pentane (anhydrous) & Sigma-Aldrich & Type I  \\
 hexane  & Sigma-Aldrich & Type I  \\
 heptane (anhydrous) &  Sigma-Aldrich & Type I    \\
 nonane  &  Sigma-Aldrich & Type I    \\
 decane  &  Sigma-Aldrich & Type I    \\
 dodecane  &  Sigma-Aldrich & Type I    \\
 hexadecane (anhydrous)  &  Sigma-Aldrich & Type I    \\
 methanol  &  Sigma-Aldrich & Type I    \\
 methanol (anhydrous) &  Sigma-Aldrich & Type I    \\
 ethanol  &  Sigma-Aldrich & Type I    \\
 ethanol (anhydrous)  &  Fisher Scientific & Type I    \\
 1-propanol (anhydrous)  &  Sigma-Aldrich & Type I   \\
 1-butanol (anhydrous)  &  Sigma-Aldrich & Type I    \\
 1-pentanol  &  Sigma-Aldrich & Type I    \\
 isopropanol  &  Sigma-Aldrich & Type I    \\
 tert-butanol  &  Sigma-Aldrich & Type I    \\
 acetone  &  Sigma-Aldrich & Type I    \\
 acetonitrile (anhydrous) &  Sigma-Aldrich & Type I   \\
 dimethylsulfoxide  &  Sigma-Aldrich & Type I    \\
 dimethylformamide  &  Sigma-Aldrich & Type I   \\
 cyclohexane  &  Sigma-Aldrich & Type I    \\
 chloroform  &  Sigma-Aldrich & Type I    \\
 dichloromethane  &  Sigma-Aldrich & Type I    \\
 chloroform  & Sigma-Aldrich & Type I    \\
 \hline
 ethylene glycol  &  Sigma-Aldrich & Type II    \\
 glycerol  & Sigma-Aldrich & Type II    \\
 benzyl alcohol  & Sigma-Aldrich & Type II    \\
 propane-1,3-diol  &  Sigma-Aldrich & Type II    \\
 \hline
 deionized water  &  MilliQ & Type III    \\
 deuterium oxide  &  Sigma-Aldrich & Type III    \\
 hydrogen peroxide  &  Sigma-Aldrich & Type III    \\
\hline
\end{tabularx}
\vspace{1 cm}

\subsection*{Supplementary Table 2: List of symbols}
\addcontentsline{toc}{subsection}{Supplementary Table 2: List of symbols}
\textbf{Super-resolution}\\
\begin{tabularx}{\textwidth}[!hb] { 
  | >{\centering\arraybackslash}X 
  | >{\centering\arraybackslash}X | }
 \hline
$N_\text{loc}$ & Number of photons of a localization event\\
$\sigma_\text{PSF}$ & Point spread function radial standard deviation\\
$\sigma_\text{loc} \approx \sigma_\text{PSF}/\sqrt{N_\text{loc}}$ & Uncertainty of localization\\

$I_\text{crystal}=\sum_\text{frame} N_\text{loc}/S\Delta t$ & Crystal brightness\\

\hline
\end{tabularx}\\

\textbf{Tracking}\\
\begin{tabularx}{\textwidth}[!hb] { 
  | >{\centering\arraybackslash}X 
  | >{\centering\arraybackslash}X | }
 \hline
$x$ & 1D displacement coordinate\\
$t$ & Time coordinate\\
$\tau$ & Lag time\\
$P(\,)$ & Probability notation\\
$PDF(x,\tau)=P\Big(X(t+\tau)-X(t)=x\Big)$ & 1D displacement probability density function\\
$D$  & Diffusion coefficient\\
$T_\text{res}^D$ & Residence time at a single defect \\
$T_\text{res}^T$ & Residence time for a whole trajectory\\
$\tau_\text{res}^D$ & Long exponential time constant in the distribution of $T_\text{res}^D$\\
$\tau_\text{res}^T$ & Long exponential time constant in the distribution of $T_\text{res}^T$\\

\hline
\end{tabularx}\\

\textbf{Liquids - nanoslits}\\
\begin{tabularx}{\textwidth}[!hb] { 
  | >{\centering\arraybackslash}X 
  | >{\centering\arraybackslash}X | }
 \hline
 $h$ & Nanoslit height\\
 $w$ & Nanoslit width\\
$\epsilon_\text{liq}$ & Bulk static dielectric constant\\
$\epsilon_\text{conf}$ & Confined static dielectric constant\\
$\epsilon_\text{wall}$ & Mica top wall static dielectric constant\\
$\mu_D$ & Defect dipole moment\\
$\mu_S$ & Solvent molecule dipole moment\\

$\ell_\text{dip}$ & Range of dipolar interactions\\
\hline
\end{tabularx}

\clearpage
\section*{Supplementary References}
\addcontentsline{toc}{section}{Supplementary References}
1. Secchi, E., Marbach, S., Niguès, A., Stein, D., Siria, A. \& Bocquet, L. Massive radius-dependent flow slippage in carbon
nanotubes. \textit{Nature} \textbf{537}, 210–213 (2016).\\
2. Seal, A. \& Govind Rajan, A. Modulating water slip using atomic-scale defects: Friction on realistic hexagonal boron
nitride surfaces. \textit{Nano Lett.} \textbf{21}, 8008–8016 (2021).\\
3. Wong, D. et al. Characterization and manipulation of individual defects in insulating hexagonal boron nitride using
scanning tunnelling microscopy. \textit{Nat. Nanotechnol.} \textbf{10}, 949–953 (2015).\\
4. Demchenko, A. P. Photobleaching of organic fluorophores: Quantitative characterization, mechanisms, protection. \textit{Methods
Appl. Fluoresc.} \textbf{8}, 022001 (2020).\\
5. Magonov, S. N., Elings, V. \& Whangbo, M. H. Phase imaging and stiffness in tapping-mode atomic force microscopy.
Surf. Sci. \textbf{375}, L385–L391 (1997).\\
6. Pang, G. K. H., Baba-Kishi, K. Z. \& Patel, A. Topographic and phase-contrast imaging in atomic force microscopy.
\textit{Ultramicroscopy} \textbf{81}, 35–40 (2000).\\
7. Caldwell, J. D., Aharonovich, I., Cassabois, G., Edgar, J. H., Gil, B. \& Basov, D. N. Photonics with hexagonal boron
nitride. \textit{Nat. Rev. Mater.} \textbf{4}, 552–567 (2019).\\
8. Geick, R., Perry, C. \& Rupprecht, G. Normal modes in hexagonal boron nitride. \textit{Phys. Rev.} \textbf{146}, 543 (1966).\\
9. Reich, S., Ferrari, A., Arenal, R., Loiseau, A., Bello, I. \& Robertson, J. Resonant raman scattering in cubic and hexagonal
boron nitride. \textit{Phys. Rev. B} \textbf{71}, 205201 (2005).\\
10. Serrano, J., Bosak, A., Arenal, R., Krisch, M., Watanabe, K., Taniguchi, T., Kanda, H., Rubio, A. \& Wirtz, L. Vibrational
properties of hexagonal boron nitride: inelastic X-ray scattering and ab initio calculations. \textit{Phys. Rev. Lett.} \textbf{98}, 095503
(2007).\\
11. Chejanovsky, N. et al. Structural attributes and photodynamics of visible spectrum quantum emitters in hexagonal boron
nitride. \textit{Nano Lett.} \textbf{16}, 7037–7045 (2016).\\
12. Martínez, L., Pelini, T., Waselowski, V., Maze, J., Gil, B., Cassabois, G. \& Jacques, V. Efficient single photon emission
from a high-purity hexagonal boron nitride crystal. Phys. Rev. B \textbf{94}, 121405 (2016).\\
13. Vuong, T., Cassabois, G., Valvin, P., Ouerghi, A., Chassagneux, Y., Voisin, C. \& Gil, B. Phonon-photon mapping in a color
center in hexagonal boron nitride. \textit{Phys. Rev. Lett.} \textbf{117}, 097402 (2016).\\
14. Sainsbury, T., Satti, A., May, P., Wang, Z., McGovern, I., Gun’ko, Y. K. \& Coleman, J. Oxygen radical functionalization of
boron nitride nanosheets. \textit{J. Am. Chem. Soc.} \textbf{134}, 18758–18771 (2012).\\
15. Jiang, T., Le, D., Rawal, T. B. \& Rahman, T. S. Syngas molecules as probes for defects in 2D hexagonal boron nitride:
their adsorption and vibrations. \textit{Phys. Chem. Chem. Phys.} \textbf{23}, 7988–8001 (2021).\\
16. Romanos, J. et al. Infrared study of boron–carbon chemical bonds in boron-doped activated carbon. \textit{Carbon} \textbf{54}, 208–214
(2013).\\
17. Lvova, N. \& Ananina, O. Y. Theoretical study of the adsorption properties of porous boron nitride nanosheets.\textit{ Comput.
Mater. Sci.} \textbf{115}, 11–17 (2016).\\
18. Piao, Y., Meany, B., Powell, L. R., Valley, N., Kwon, H., Schatz, G. C. \& Wang, Y. Brightening of carbon nanotube
photoluminescence through the incorporation of sp3 defects. \textit{Nat. Chem.} \textbf{5}, 840–845 (2013).\\
19. Nigam, S. \& Rutan, S. Principles and applications of solvatochromism. \textit{Appl. Spectrosc.} \textbf{55}, 362A–370A (2001).\\
20. Yomogida, Y., Sato, Y., Nozaki, R., Mishina, T. \& Nakahara, J. Dielectric study of normal alcohols with THz time-domain
spectroscopy. \textit{J. Mol. Liq.} \textbf{154}, 31–35 (2010).\\

\end{document}